\documentclass[12pt]{article}

\input epsf

\renewcommand{\arraystretch}{2.0} 

\def\dfrac#1#2{{\displaystyle {#1 \over #2}}}
\def\dsum{\mathop{\displaystyle \sum }}

\def\simge{\mathrel{\rlap{\raise 0.511ex \hbox{$>$}}{\lower 0.511ex \hbox{$\sim$}}}}
\def\simle{\mathrel{\rlap{\raise 0.511ex \hbox{$<$}}{\lower 0.511ex \hbox{$\sim$}}}} 
\def\slash#1{\setbox0=\hbox{$#1$}\dimen0=\wd0                    
      \setbox1=\hbox{/} \dimen1=\wd1 \ifdim\dimen0>\dimen1
      \rlap{\hbox to \dimen0{\hfil/\hfil}} #1                        \else                                       
      \rlap{\hbox to \dimen1{\hfil$#1$\hfil}}
      /   \fi}                                         

\newcommand{\be}{\begin{equation}}
\newcommand{\ee}{\end{equation}}
\newcommand{\bea}{\begin{eqnarray}}
\newcommand{\eea}{\end{eqnarray}}
\newcommand{\nn}{\nonumber}
\newcommand{\as}{\alpha_{ s}}

\newcommand{\Heff}{{\cal H}_{ eff}}
\newcommand{\DB}{\Delta B}

\newcommand{\msb}{\mbox{NDR-}\overline{\rm{MS}}}

\newcommand{\ri}{\mbox{RI-MOM}}
\newcommand{\ep}{\varepsilon}
\begin{document}

\begin{titlepage}
\begin{flushright}
RM3-TH/01-7 \\
ROME1-1326/01
\end{flushright}
\vskip 2.4cm
\begin{center}
{\Large \bf Next-to-Leading Order QCD Corrections \\ to Spectator Effects 
\\  \vspace{0.2cm} in Lifetimes of Beauty Hadrons}

\vskip1.3cm 
{\large\bf M.~Ciuchini$^a$, E.~Franco$^b$, V.~Lubicz$^a$ and F.~Mescia$^b$}\\

\vspace{1.cm}
{\normalsize {\sl 
$^a$ Dip. di Fisica, Univ. di Roma Tre and INFN,
Sezione di Roma III, \\
Via della Vasca Navale 84, I-00146 Rome, Italy \\
\vspace{.25cm}
$^b$ Dip. di Fisica, Univ. di Roma ``La Sapienza" and INFN,
Sezione di Roma,\\ P.le A. Moro 2, I-00185 Rome, Italy }}\\
\vskip1.5cm
%{\large\bf Abstract:\\[10pt]} \parbox[t]{\textwidth}{ 
\abstract{
Theoretical predictions of beauty hadron lifetimes, based on the heavy quark
expansion up to and including order $1/m_b^2$, do not to reproduce the
experimental measurements of the lifetime ratios $\tau(B^+)/\tau(B_d)$ and
$\tau(\Lambda_b)/\tau(B_d)$. Large corrections to these predictions come from
phase-space enhanced $1/m_b^3$ contributions, i.e. hard spectator effects. In 
this paper we calculate the next-to-leading order QCD corrections to the Wilson 
coefficients of the local operators appearing at ${\cal O}(1/m_b^3)$.
We find that these corrections improve the agreement with the experimental data.
The lifetime ratio of charged to neutral $B$-mesons, $\tau(B^+)/\tau(B_d)$,
turns out to be in very good agreement with the corresponding measurement, whereas
for $\tau(B_s)/\tau(B_d)$ and $\tau(\Lambda_b)/\tau(B_d)$ there is a residual difference
at the 1$\sigma$ level. We discuss, however, why the theoretical predictions are less accurate
in the latter cases.}
\end{center}
\vspace*{1.cm}
%%{\footnotesize{\tt PACS numbers: 12.15Ff,\ 11.15.Ha,\ 12.38.Gc.}}
\end{titlepage}

\setcounter{footnote}{0}
\setcounter{equation}{0}

%%%%%%%%%%%%%%%%%%%%%%%%%%%%%%%%%%%%%%%%
\section{Introduction}
\label{sec:intro}
%%%%%%%%%%%%%%%%%%%%%%%%%%%%%%%%%%%%%%%%
The calculation of inclusive decay rates of hadrons containing a $b$ quark can
be performed by computing the amplitude through an Operator Product Expansion (OPE)
in powers of $\Lambda_{QCD}/m_b$~\cite{Khoze:1987fa,ope}. This allows a
quantitative evaluation of the hadronic decay rates under the assumption that
properly smeared partonic amplitudes can replace the hadronic ones. This 
assumption is usually referred to as the ``quark-hadron duality'', which in the 
case of total rates is required to hold in its {\it local} 
formulation~\cite{dua1}-\cite{dua3}, namely the smearing is provided by the sum over 
exclusive states.

Within this theoretical framework, and up to terms of ${\cal O}(1/m_b^2)$, only
the $b$ quark  enters the short-distance weak decay, while the light quarks in
the hadron interact through soft gluons only. In particular, the leading term
in the expansion reproduces the results of the old spectator model, thus
providing a theoretical basis to this model. Quantitatively, however, the
contributions to the lifetimes ratios of the first two terms in the OPE are too
small. Neglecting terms of ${\cal O}(1/m_b^3)$, one finds
\be
\label{eq:rth}
\frac{\tau(B^+)}{\tau(B_d)}=1.00
\, , \qquad
\frac{\tau(B_s)}{\tau(B_d)}=1.00
\, , \qquad
\frac{\tau(\Lambda_b)}{\tau(B_d)}= 0.98(1)
\, ,
\ee
to be compared with the experimental measurements~\cite{pdg}
\be
\label{eq:rexp}
\frac{\tau(B^+)}{\tau(B_d)}=1.07 \pm 0.02 \, , \qquad
\frac{\tau(B_s)}{\tau(B_d)}=0.95 \pm 0.04 \, , \qquad
\frac{\tau(\Lambda_b)}{\tau(B_d)}=0.80 \pm 0.06\, .
\ee
Given the discrepancies, it is important to understand whether we are facing a 
signal of quark-hadron duality violation or, instead, the inclusion of
higher-order terms in $\Lambda_{QCD}/m_b$ and $\alpha_s$ is sufficient to
reproduce the data.

Spectator contributions, which appear at ${\cal O}(1/m_b^3)$ in the OPE, can
explain the lifetime differences, as they distinguish the light-quark
content of the hadrons. As observed by Neubert and Sachrajda~\cite{NS}, these
effects, although suppressed by an additional power of $1/m_b$, are enhanced
with respect to leading contributions by a phase-space factor $16\pi^2$, being
$2\to 2$ processes instead of $1\to 3$ decays.
The calculation of spectator effects at the leading order (LO) in QCD has been
done few years ago~\cite{NS,primisp} and a phenomenological analysis has been
presented in ref.~\cite{NS}. By using the recent lattice calculations of the
relevant four-fermion operator matrix elements~\cite{DiPierro98}-\cite{APE},
we obtain the LO predictions
\be
\label{eq:rlo}
\left. \frac{\tau(B^+)}{\tau(B_d)}\, \right|_{\rm LO}\!\!\!\!\! =
1.01 \pm 0.03 \, ,  \qquad
\left. \frac{\tau(B_s)}{\tau(B_d)}\, \right|_{\rm LO}\!\!\!\!\! =
1.00 \pm 0.01 \, ,  \qquad
\left. \frac{\tau(\Lambda_b)}{\tau(B_d)}\, \right|_{\rm LO}\!\!\!\!\! =
0.93 \pm 0.04\, ,
\ee
in agreement with previous estimates. The errors include the uncertainty
due to the variation of
the renormalization scale $\mu$ between $m_b/2$ and $2m_b$.
For the ratio $\tau(\Lambda_b)/\tau(B_d)$, the inclusion of spectator
contributions reduces the discrepancy between theoretical and experimental
determinations while, for $B$ mesons, at the LO, no significant improvement is
seen.

In this paper we compute the next-to-leading order (NLO) QCD corrections to the
coefficient functions of the valence operators contributing to spectator
effects. Our calculation has been performed in the limit of vanishing charm
quark mass, which means that corrections of the order of $\alpha_s m_c^2/m_b^2$
have been neglected. The main motivations to improve the leading-order results
are
\begin{itemize}
\item
reducing the large renormalization-scale dependence which appears at the
LO~\cite{NS};
\item
properly taking into account in the matching the renormalization-scheme
dependence of the Wilson coefficients and of the operator matrix elements;
\item
increasing the accuracy of the theoretical predictions by evaluating ${\cal O}(\as)$
corrections which are potentially large.
\end{itemize}

Using the results of the NLO calculation, and the available lattice
determinations of the hadronic matrix elements~\cite{DiPierro98}-\cite{APE}, we
have also performed in this paper a phenomenological analysis of the lifetime
ratios. We find that the inclusion of NLO QCD corrections to spectator effects
improves the agreement between theoretical predictions of lifetime ratios and
their measured values. We obtain the estimates
\be
\label{eq:nlores}
\left. \frac{\tau(B^+)}{\tau(B_d)}\, \right|_{\rm NLO}\!\!\!\!\! =
1.07 \pm 0.03 \, ,  \qquad
\left. \frac{\tau(B_s)}{\tau(B_d)}\, \right|_{\rm NLO}\!\!\!\!\! =
1.00 \pm 0.01 \, ,  \qquad
\left. \frac{\tau(\Lambda_b)}{\tau(B_d)}\, \right|_{\rm NLO}\!\!\!\!\! =
0.89 \pm 0.05\, ,
\ee
to be compared with the experimental determinations given in
eq.~(\ref{eq:rexp}). The lifetime ratio of charged to neutral $B$-mesons
turns out to be in very good agreement with the experimental data. In the case
of the $B_s$ meson and the $\Lambda_b$ baryon the agreement is at the 1$\sigma$
level. We should also mention, however, that, in the baryon case, the lattice
evaluation of the relevant matrix elements is still
preliminary~\cite{DiPierro:1999tb}.

An important check of our perturbative calculation is provided by the
cancellation of the infrared (IR) divergences in the expressions of the
coefficient functions, in spite these divergences appear in the individual
amplitudes. Their presence provides an example of violation of the
Bloch-Nordsieck theorem in non-abelian gauge theories. We have also checked
that our results are explicitly gauge invariant and have the correct
ultraviolet (UV) renormalization-scale dependence as predicted by the known
anomalous dimensions of the relevant operators.

The OPE of the lifetime ratios is expressed in terms of local operators defined
in the Heavy Quark Effective Theory (HQET). The non-perturbative lattice
determination of the hadronic matrix elements has been performed, in some
cases, in terms of operators defined in QCD. In order to combine these
determinations with the results obtained for the Wilson coefficients,
consistently at the NLO, a matching between QCD and the HQET operators must be
performed. In this paper, we have computed this matching, at ${\cal O}(\as)$,
in the case of the flavour non-singlet $\DB=0$ four fermion operators, for
which the lattice results are available so far.

The NLO predictions of the lifetime ratios may still be improved in several
ways. These improvements concern both the perturbative and the non-perturbative
part of the calculation, and are:
\begin{itemize}
\item[-] the charm quark mass corrections, of the order
$\as m_c^2/m_b^2$, to the Wilson coefficients are still unknown. Such a
calculation is in progress;

\item[-] the NLO coefficient functions of the current-current
operators containing the charm quark field and the penguin operator (see
eqs.~(\ref{eq:opeff}) and (\ref{eq:openg})) have not been computed yet. The
non-perturbative determination of the corresponding matrix elements is also
missing. These contributions do not affect the theoretical determination of
$\tau(B^+)/\tau(B_d)$ and are an $SU(3)$-breaking effect for $\tau(B_s)/\tau(B_d)$.
In the case of $\tau(\Lambda_b)/\tau(B_d)$, however, their effects may not be
negligible;

\item[-] the NLO anomalous dimension of the relevant operators
in the HQET is still unknown. For this reason, present lattice calculations
of the $\Lambda_b$ matrix elements~\cite{DiPierro:1999tb,DiPierroproc}
only achieve a LO accuracy. For $B$ mesons, a complete NLO determination
has been performed on the lattice by using operators defined in QCD~\cite{APE};

\item[-] in the lattice determination of the hadronic matrix elements,
penguin contractions (i.e. eye diagrams), which contribute in the case of flavour non-singlet
operators, have not been computed. These contributions cancel in the evaluation
of the ratio $\tau(B^+)/\tau(B_d)$ but affect $\tau(\Lambda_b)/\tau(B_d)$ and
$\tau(B_s)/\tau(B_d)$, the latter through $SU(3)$-breaking effects.
\end{itemize}
The above discussion shows that, while the theoretical determination of
$\tau(B^+)/\tau(B_d)$ can be considered as rather accurate, in the case of
$\tau(B_s)/\tau(B_d)$ and $\tau(\Lambda_b)/\tau(B_d)$ some ingredients are still
missing, and their determination may further improve the agreement with the
experimental data.

The plan of this paper is the following. In sect.~\ref{sec:formulae} we give
the relevant formulae for the decay rate of $b$-hadrons. The details of the
calculation are given in sect.~\ref{sec:details}, while our results are
summarized in sect.~\ref{sec:result}. In sect.~\ref{sec:match} we present the
results of the matching between QCD and HQET operators for the flavour
non-singlet sector. Finally, the phenomenological analysis is presented in 
sect.~\ref{sec:phenom}.

%%%%%%%%%%%%%%%%%%%%%%%%%%%%%%%%%%%%%%%%%%%%%%%%%%%%%%%%%%%%
\section{The inclusive decay width of beauty hadrons}
\label{sec:formulae}
%%%%%%%%%%%%%%%%%%%%%%%%%%%%%%%%%%%%%%%%%%%%%%%%%%%%%%%%%%%%
Using the optical theorem, the inclusive decay width of a hadron $H_b$
containing a $b$ quark can be written as the imaginary part of the forward
matrix element of the transition operator ${\cal T}$
\be
\Gamma(H_b \to X) =
\frac{1}{M_{H_b}} {\mathrm Im} \langle H_b \vert {\cal T} \vert H_b \rangle =
\frac{1}{2 M_{H_b}} \langle H_b \vert \widehat \Gamma \vert H_b
\rangle\, ,
\label{eq:master}
\ee
where ${\cal T}$ is given by
\be 
{\cal T} = i \int d^4x \; T \left( \Heff^{\DB=1}(x) \Heff^{\DB=1}(0) \right)\, .
\label{eq:T}
\ee
The operator $\Heff^{\DB=1}$ is the effective weak hamiltonian which describes 
$\DB=1$ transitions. It has the form
\bea
\label{eq:hdb1}
&&\Heff^{\DB=1} =
\frac{G_F}{\sqrt{2}} \Bigg\{ \Bigg[\Bigg( V^\ast_{cb} V_{us} \left( C_1 Q_1 + 
C_2 Q_2 \right) + V^\ast_{cb} V_{cs} \left( C_1 Q^c_1 + C_2 Q^c_2 \right) + 
\left( c \leftrightarrow u \right) \Bigg) -  \nn \\
&&\qquad V^\ast_{tb} V_{ts} \left( \sum_{i=3}^{6} C_i Q_i + C_{8G} Q_{8G} 
\right) + \left[ s \to d \right] \Bigg]+\dsum_{l=e,\mu,\tau}\left(V^\ast_{cb} 
Q_l^c +V^\ast_{ub} Q_l^u\right) \Bigg\} + h.c.\,,
\eea 
where $C_i$ are the Wilson coefficients, known at the NLO in perturbation
theory~\cite{nlodb1a}-\cite{nlodb1c}, and the operators $Q_i$ are defined as
\be
\begin{array}{ll}
Q_1= (\bar b_i c_j)_{V-A} (\bar u_j s_i)_{V-A}\,, \qquad \qquad
& Q_2= (\bar b_i c_i)_{V-A} (\bar u_j s_j)_{V-A}\,, \\
Q^c_1= (\bar b_i c_j)_{V-A} (\bar c_j s_i)_{V-A}\,, \qquad \qquad
& Q^c_2= (\bar b_i c_i)_{V-A} (\bar c_j s_j)_{V-A}\,, \\
Q_3= (\bar b_i s_i)_{V-A} \dsum\limits_{q} (\bar q_j q_j)_{V-A}\,, \qquad \qquad
& Q_4= (\bar b_i s_j)_{V-A} \dsum\limits_{q} (\bar q_j q_i)_{V-A}\,, \\
Q_5= (\bar b_i s_i)_{V-A} \dsum\limits_{q} (\bar q_j q_j)_{V+A}\,, \qquad \qquad
& Q_6= (\bar b_i s_j)_{V-A} \dsum\limits_{q} (\bar q_j q_i)_{V+A}\,, \\
Q_{8G}= \dfrac{g_s}{8\pi^2} m_b \bar b_i \sigma^{\mu\nu} 
\left(1-\gamma^5\right) t^{a}_{ij} s_j G^{a}_{\mu\nu} & Q_l^q=(\bar b_i
q_i)_{V-A}(\bar\nu_l l)_{V-A}\,.
\end{array}
\label{eq:operatori}
\ee
Here and in the following we use the notation $(\bar q q)_{V\pm A}=\bar q 
\gamma_\mu(1\pm\gamma_5) q$. A sum over colour indices is always understood.
We also neglect Cabibbo-suppressed terms in eq.~(\ref{eq:hdb1}), so that the 
effective $\DB=1$ hamiltonian reduces to
\bea
\Heff^{\DB=1}&=&
\frac{G_F}{\sqrt{2}} V^\ast_{cb} \Bigg\{ 
C_1 \left[(\bar b_i c_j)_{V-A} (\bar u_j d_i)_{V-A} + 
(\bar b_i c_j)_{V-A} (\bar c_j s_i)_{V-A} \right] + \nn \\
&& \qquad\quad~\, C_2 \left[(\bar b_i c_i)_{V-A} (\bar u_j d_j)_{V-A} +
 (\bar b_i c_i)_{V-A} (\bar c_j s_j)_{V-A} \right] +\\ 
&&\qquad\quad~\sum_{i=3}^{6} C_i Q_i + C_{8G} Q_{8G} +
\dsum_{l=e,\mu,\tau} Q_l^c\,\Bigg\} + h.c. \, .\nn
\eea 
In the double insertion of the $\DB=1$ effective hamiltonian, this corresponds
to neglecting terms suppressed by $m_c^2/m_b^2\sin^2 \theta_c$ with respect
to the dominant ones.

\begin{figure} [t]
\begin{center}
%\vspace*{-.55cm}
\epsfxsize=12cm
\epsfbox{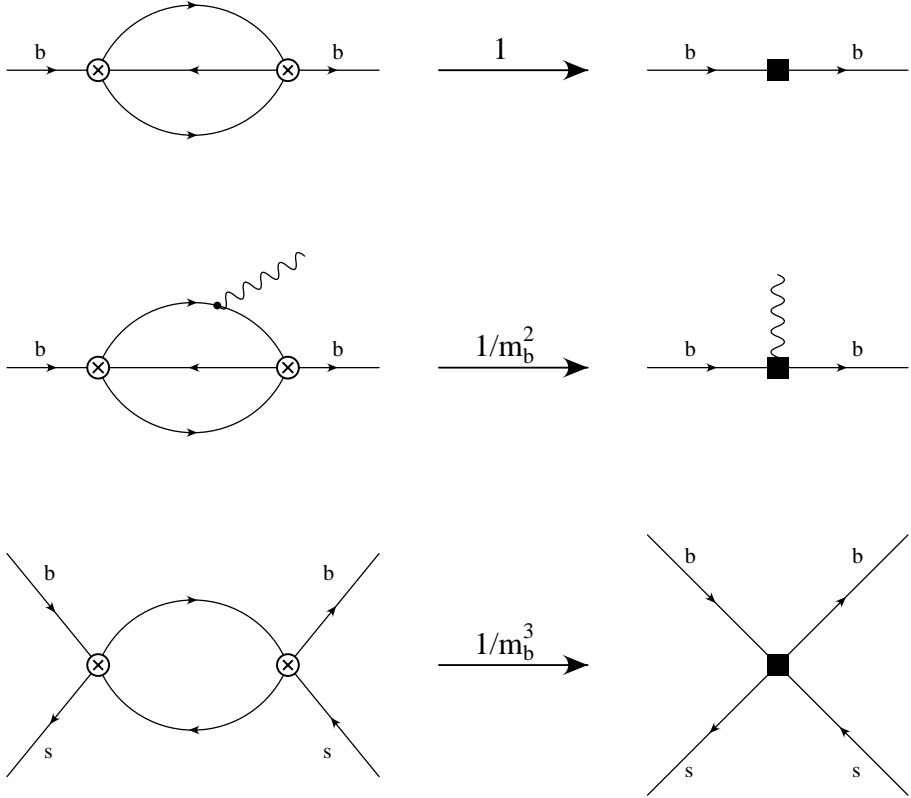}
%\vspace*{-1.3cm}
\end{center}
\caption{\it Examples of LO contributions to the transition operator ${\cal T}$
(left) and to the corresponding local operator $\widehat \Gamma$ (right). The
crossed circles represent the insertions of the $\DB=1$ effective hamiltonian.
The black squares represent the insertion of a $\DB=0$ operator.}
\label{fig:tree}
\end{figure}

Because of the large mass of the $b$ quark, it is possible to construct an OPE
for the transition operator ${\cal T}$ of eq.~(\ref{eq:master}), which results
in a sum of local operators of increasing dimension~\cite{Khoze:1987fa,ope}.
We include in
this expansion terms up to ${\cal O}(1/m_b^2)$ plus those $1/m_b^3$
corrections that come from spectator effects and are enhanced by the phase 
space. The resulting expression for the local $\DB=0$ operator $\widehat 
\Gamma$, of eq.~(\ref{eq:master}), is
\be
\widehat \Gamma=\frac{G_F^2 |V_{cb}|^2 m_b^5}{192 \pi^3}\left[
c^{(3)} \bar b b +
c^{(5)} \frac{g_s}{m_b^2}\bar b\sigma_{\mu\nu}G^{\mu\nu} b +
\frac{96\pi^2}{m_b^3} \dsum_{k} c^{(6)}_k O_k^{(6)} \right]\, ,
\label{eq:gamma}
\ee
where the $O_k^{(6)}$ are a set of four-fermion dimension-six operators to
be specified below. These operators represent the contribution of hard 
spectator effects. At the lowest order in QCD, the diagrams 
entering the calculation of $\widehat \Gamma$ are shown in 
fig.~\ref{fig:tree}. 
%________________________________________________________________
%________________________________________________________________

We notice that, in general, the OPE of $\widehat \Gamma$ cannot be expressed
in terms of local operators defined in QCD. The reason is that, in full QCD,
renormalized operators mix with operators of lower dimension
with coefficients proportional to powers of the $b$-quark mass. In this
way, the dimensional ordering of the OPE is lost. In order to implement the
expansion, the matrix elements of the local operators should be cut-off at a
scale smaller than the $b$-quark mass. This is naturally realized by defining
the operators in the HQET.

From eqs.~(\ref{eq:master}) and (\ref{eq:gamma}), one can derive an expression
for the ratio of inclusive widths
\be
\label{eq:gratio}
\dfrac{\Gamma(H_b)}{\Gamma(H_b^\prime)}=\dfrac{M_{H_b^\prime}\langle \bar b b
\rangle_{H_b}}{M_{H_b}\langle \bar b b\rangle_{H_b^\prime}}
\left(\dfrac{1+\dfrac{g_s}{m_b^2}\dfrac{c^{(5)}}{c^{(3)}}
\dfrac{\langle\bar b\sigma_{\mu\nu}G^{\mu\nu} b\rangle_{H_b}}{\langle \bar b b
\rangle_{H_b}}
+ \dfrac{96\pi^2}{m_b^3} \dsum\limits_{k}
\dfrac{c^{(6)}_k \langle O_k\rangle_{H_b}}{c^{(3)}\langle \bar b b
\rangle_{H_b}}}{1+\dfrac{g_s}{m_b^2}\dfrac{c^{(5)}}{c^{(3)}}
\dfrac{\langle\bar b\sigma_{\mu\nu}G^{\mu\nu} b\rangle_{H_b^\prime}}{\langle
\bar b b\rangle_{H_b^\prime}}
+ \dfrac{96\pi^2}{m_b^3}\dsum\limits_{k}
\dfrac{c^{(6)}_k \langle O_k\rangle_{H_b^\prime}}{c^{(3)}\langle \bar b
b\rangle_{H_b^\prime}}}
\right)\, ,
\ee
where $\langle\cdots\rangle_{H}$ denotes the forward matrix element between
two hadronic states $H$ defined with the covariant normalization. The matrix
elements of dimension-three and dimension-five operators, appearing in
eq.~(\ref{eq:gratio}), can be expanded using the HQET
\bea
\langle \bar b b\rangle_{H_b}&=&2 M_{H_b} \left( 1-\frac{\mu_\pi^2(H_b)-
\mu_G^2(H_b)}{2 m_b^2}+{\cal O}(1/m_b^3)\right)\, ,\nn \\
g_s
\langle\bar b\sigma_{\mu\nu}G^{\mu\nu} b\rangle_{H_b}&=& 2M_{H_b}
\left(2\mu^2_G(H_b)+{\cal O}(1/m_b)\right)\, .
\label{eq:hqet}
\eea
By substituting these expansions in eq.~(\ref{eq:gratio}), we finally obtain
\bea
\label{eq:ratio}
\dfrac{\Gamma(H_b)}{\Gamma(H_b^\prime)}&=&1-\frac{\mu_\pi^2(H_b)-
\mu_\pi^2(H_b^\prime)}{2 m_b^2}+\left(\frac{1}{2}+\frac{2 c^{(5)}}{c^{(3)}}
\right) \frac{\mu_G^2(H_b)-\mu_G^2(H_b^\prime)}{m_b^2}+\nn\\
&&\quad\dfrac{96\pi^2}{m_b^3\, c^{(3)}}\dsum\limits_{k}
c^{(6)}_k \left(\frac{\langle O^{(6)}_k\rangle_{H_b}}{2M_{H_b}}-\frac{\langle
O_k^{(6)}\rangle_{H_b^\prime}}{2M_{H_b^\prime}}\right) \, ,
\eea
which is the expression used in the evaluation of the lifetime ratios of beauty
hadrons.

The Wilson coefficients $c^{(3)}$ and $c^{(5)}$ in eq.~(\ref{eq:ratio}) have
been computed at the LO in ref.~\cite{Bigi:1992su}, while the NLO corrections
to $c^{(3)}$ have been evaluated in \cite{gamma1}-\cite{gamma6}. The NLO corrections to
$c^{(5)}$ are still missing. Their numerical contribution to the lifetime
ratios, however, is expected to be negligible.

The dimension-six operators in eq.~(\ref{eq:ratio}), which express the hard
spectator contributions, are the current-current operators
\be
\begin{array}{ll}
O^q_1= (\bar b b)_{V-A} (\bar q q)_{V-A}\,, &
O^q_2= (\bar b b)_{V+A} (\bar q q)_{V-A}\,, \\
O^q_3= (\bar b t^{a} b)_{V-A} (\bar q t^{a} q)_{V-A}\,, &
O^q_4= (\bar b t^{a} b)_{V+A} (\bar q t^{a} q)_{V-A}\, ,
\end{array}
\label{eq:opeff}
\ee
with $q=u,d,s,c$, and the penguin operator
\be
O_P= (\bar b_i t^{a}_{ij} b_j)_{V}
 \dsum\limits_{q=u,d,s,c}(\bar q_k t^{a}_{kl} q_l)_{V} \, .
\label{eq:openg}
\ee
In these definitions, the symbols $b$ and $\bar b$ denote the heavy quark
fields in the HQET. Note that the operator basis in eq.~(\ref{eq:opeff})
differs from the one used in ref.~\cite{NS} by a Fierz rearrangement. Our choice
of the basis is more convenient for NLO calculations because it does not
require the introduction of Fierz-evanescent operators in the intermediate
steps of the calculation.

The coefficient functions of the current operators $O^q_k$, have been computed
at the LO in ref.~\cite{NS} for $q=u,d,s$, and in ref.~\cite{charm} for the
charm quark operator. The coefficient function of the penguin operator $O_P$
vanishes at the LO.

In this paper we have computed the NLO QCD corrections to the coefficient
functions of the operators $O^q_k$ with $q=u,d,s$. The operators containing the
charm quark fields contribute, as valence operators, only to the inclusive
decay rate of $B_c$ mesons, and their contribution to non-charmed
hadron decay rates is expected to be negligible. The calculation of the NLO
corrections to these coefficient functions, as well as the NLO calculation of
the coefficient function of the penguin operator, has not been performed yet.

%%%%%%%%%%%%%%%%%%%%%%%%%%%%%%%%%%%%%%%%%%%%%%%%%%%%%%%%%%%%
\section{Details of the perturbative calculation}
\label{sec:details}
%%%%%%%%%%%%%%%%%%%%%%%%%%%%%%%%%%%%%%%%%%%%%%%%%%%%%%%%%%%%
In this section we summarize the general formulae used in the matching
procedure, and present the details of our calculation. The matching condition
between the transition operator in eq.~(\ref{eq:T}) and the width operator in
eq.~(\ref{eq:gamma}) can be written in the form
\be
Im \, \langle {\cal T} \rangle \sim \vec c^{\,T}(\mu) \langle \vec O(\mu)
\rangle\, ,
\label{eq:match}
\ee
where $\langle \cdots \rangle$ indicates the matrix element computed between a
common pair of partonic external states. We are using the vector notation for
both coefficients and operators. The condition~(\ref{eq:match}) determines the
Wilson coefficients $\vec c$ at the matching scale $\mu$.

We expand at ${\cal O}(\as)$ the matrix elements in eq.~(\ref{eq:match}) and
express the result in terms of tree-level matrix elements $\langle \vec O
\rangle_0$
\bea
Im \, \langle {\cal T} \rangle &=& \left(
\vec T_0+\frac{\as}{4\pi} \vec T_1\right)^T  \langle
\vec O \rangle_0^{QCD}\,,\nn \\
\langle \vec O(\mu)\rangle &=&\left(1+\frac{\as}{4\pi}
\hat s \right) \langle\vec O \rangle_0^{HQET}\, ,
\label{eq:Texp}
\eea
We also consider the perturbative expansion for the Wilson coefficients,
\be
\vec c=\vec c_0+\frac{\as}{4\pi} \vec c_1\, .
\label{eq:COexp}
\ee
Up to $1/m_b$ corrections, the tree-level matrix elements of the operators in
QCD are equal to their HQET counterparts so that, by using
eqs.~(\ref{eq:match})--(\ref{eq:COexp}), we readily find the matching
conditions at the scale $\mu$
\be
\label{eq:match1}
\vec c_0=\vec T_0\,, \qquad \vec c_1=\vec T_1- \hat s^T \, \vec T_0\,.
\ee
%________________________________________________________________
\begin{figure}
%\vspace*{-.55cm}
\begin{center}
\epsfxsize=14cm
\epsfbox{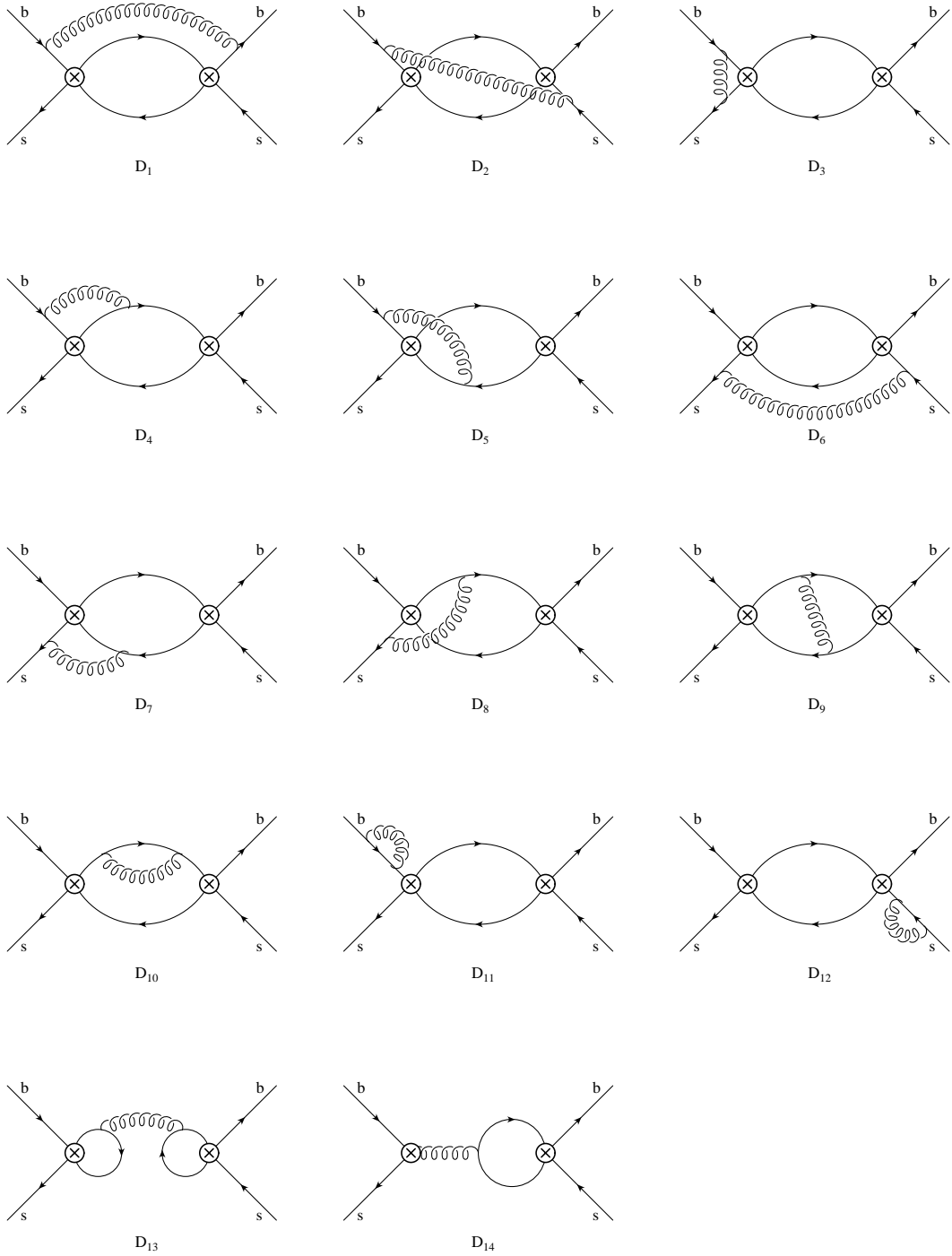}
%\vspace*{-1.3cm}
\end{center}
\caption{\it Feynman diagrams which contribute at NLO to the matrix element
of the transition operator ${\cal T}$ in the case $q=s$. In the other cases,
$q=u,\,d$, diagrams $D_{13}$ and $D_{14}$ do not contribute.
% Other diagrams, not shown in the figure, are taken into account by
% multiplying $D_2$, $D_3$, $D_4$, $D_5$, $D_7$, $D_8$, $D_{10}$, $D_{11}$,
% $D_{12}$, $D_{13}$, $D_{14}$ by a factor of 2.
}
\label{fig:fddb1}
\end{figure}
%________________________________________________________________

%________________________________________________________________
\begin{figure} [t]
%\vspace*{-.55cm}
\begin{center}
\epsfxsize=12cm
\epsfbox{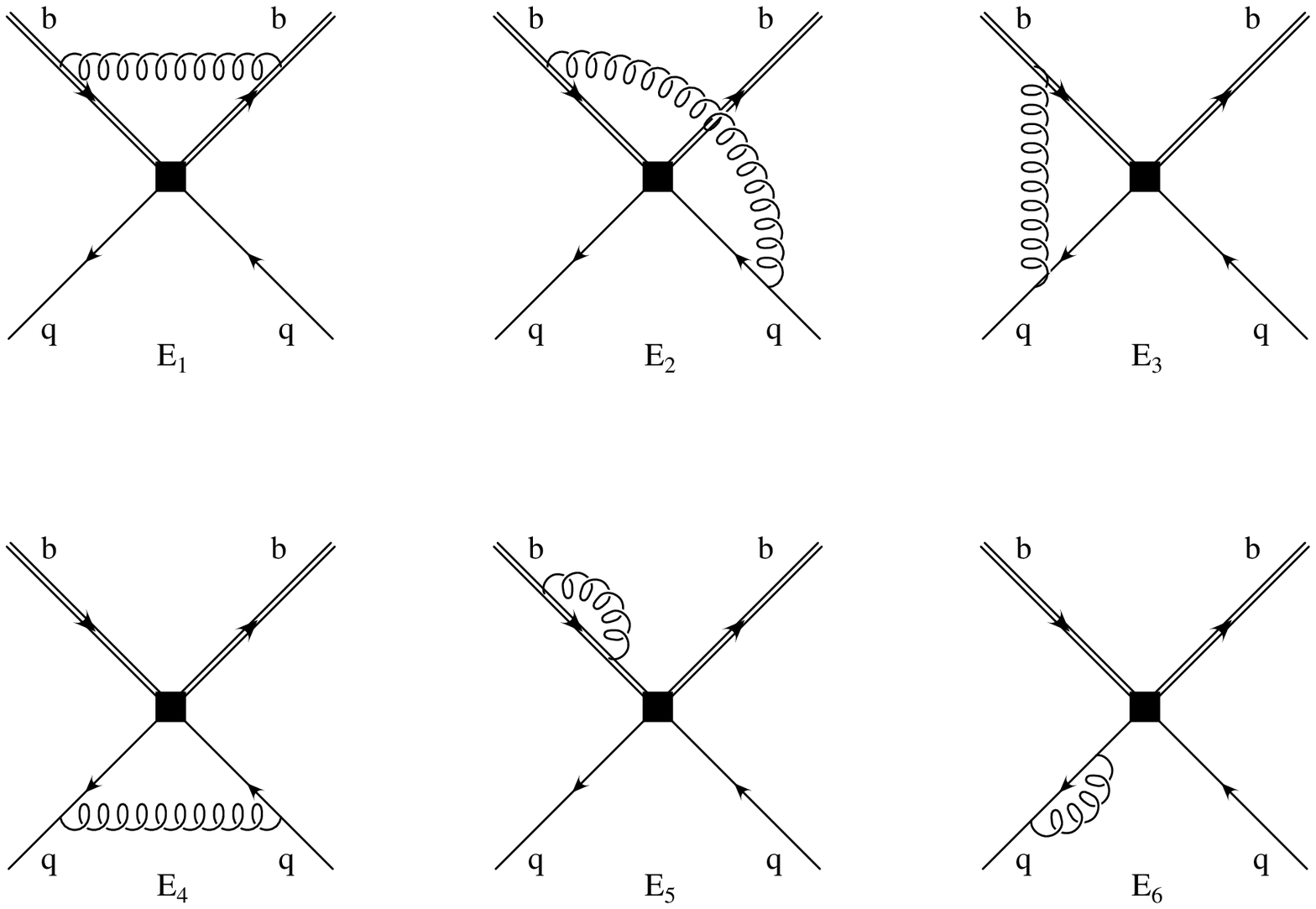}
%\vspace*{-1.3cm}
\end{center}
\caption{\it Feynman diagrams which contribute at NLO to the matrix
element of the $\DB=0$ operators entering the width operator $\widehat \Gamma$.}
\label{fig:diaeff}
%\vspace*{-.55cm}
\end{figure}
%________________________________________________________________

In order to obtain $\vec T_1$ and $\hat s$, we have first computed in QCD the
imaginary part of the diagrams shown in fig.~\ref{fig:fddb1} ({\it full}
theory) and then, in the HQET, the diagrams shown in fig.~\ref{fig:diaeff}
({\it effective} theory). The external quark states have been taken on-shell
and all quark masses, except $m_b$, have been neglected. More specifically, we
have chosen the heavy quark momenta $p_b^2=m_b^2$ in QCD and $k_b=0$ in the
HQET, and $p_q=0$ for all light quarks. In this way, we automatically retain
the leading term in the $1/m_b$ expansion. We have performed the calculation
in a generic covariant gauge, in order to check the gauge independence of the
final results. Two-loop integrals have been reduced to a set of known (massless
$p$- and massive tadpole) integrals using the recurrence relation
technique~\cite{integrals1}-\cite{integrals3}. Equations of motion have
been used to reduce the number of independent operators. The $\DB=0$ operators
in the HQET have been renormalized in the $\msb$ scheme defined in
details in ref.~\cite{reyes}.

Some diagrams, both in the full and in the effective theory,
are plagued by IR divergences. These are diagrams where a soft gluon is
exchanged between the external legs ($D_1$, $D_2$, $D_3$, $D_6$ in
fig.~\ref{fig:fddb1} and all the diagrams in
fig.~\ref{fig:diaeff})~\footnote{~We notice that the introduction of non-vanishing
masses and momenta for the external light quarks is not sufficient to eliminate the IR
divergences.}.
The divergences do not cancel in the final partonic amplitudes, and provide an
example of violation of the Bloch-Nordsieck theorem in non-abelian gauge
theories~\cite{Bloch:1937pw}--\cite{Di'Lieto:1980dt}. In agreement with this theorem, we explicitly
verified that the abelian combination of the diagrams does not contain IR
divergences. They only appear when the colour structure is taken into account.
According to the KLN theorem~\cite{kln1,kln2}, IR singularities
cancel when the contribution of soft gluons in the initial state is included in
the amplitudes. Even if the amplitudes in the full and effective
theories are IR divergent, we expect the IR poles to cancel in the matching,
since the Wilson coefficients must be insensitive to soft physics. We have
explicitly checked that, in the computation of the coefficient functions
$c^{(6)}$, this cancellation takes place.

%________________________________________________________________
\begin{figure} [t]
%\vspace*{-.55cm}
\begin{center}
\epsfxsize=14cm
\epsfbox{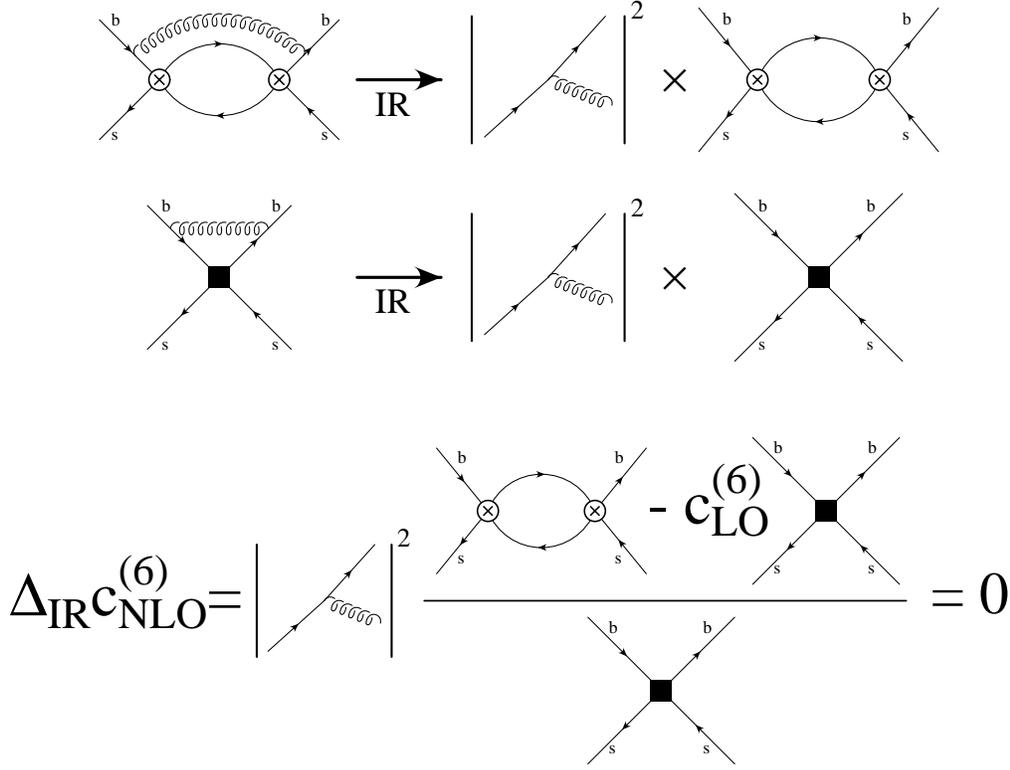}
%\vspace*{-1.3cm}
\end{center}
\caption{\it Example of the mechanism that guarantees the vanishing
of IR contributions to the coefficient functions. Only the
contributions from diagrams $D_1$ and $E_1$ are shown. In the presence of
dimensionally-regularized IR divergences, LO matching up to and including terms
of ${\cal O}(\varepsilon)$ is required.
}
\label{fig:ir}
\end{figure}
%________________________________________________________________
We use $D$-dimensional regularization with anticommuting $\gamma_5$ (NDR) to
regularize both UV and IR divergences.
The presence of dimensionally-regularized IR divergences introduces
subtleties in the matching procedure. Usually, the matching can be performed by
only considering the four dimensional operator basis, since renormalized
evanescent operators do not give contributions to the physical amplitudes. In
the present case, however, IR poles in $\varepsilon=(4-D)/2$ promote the ${\cal
O}(\varepsilon)$ contribution from evanescent operators to finite terms in the
$D\to 4$ limit. This contribution has to be taken into account and the matching
procedure must be consistently performed at ${\cal O}(\varepsilon)$. To
illustrate this point in further details, we provide the example in
fig.~\ref{fig:ir}. Let us consider the IR-divergent parts of the diagrams
$D_1$ and $E_1$. These divergences are not expected to give contribution to the
Wilson coefficients, since the coefficients should only account for the
UV behaviour of the effective theory. This condition is in fact guaranteed by
the factorization of IR divergences and by the LO matching condition. Indeed,
the numerator entering the definition of $\Delta_{IR} c^{(6)}_{\rm NLO}$ in
fig.~\ref{fig:ir}, which represents the IR contribution to the coefficient
function $c^{(6)}$ at the NLO, vanishes just because of the LO matching
condition. This happens provided that IR divergences are regularized in the
same way in the full and in the effective theory, independently of the specific
choice of the IR regulator. In this way, soft physics does not enter the
coefficient functions. If, however, IR
divergences are dimensionally regularized, the condition $\Delta_{IR} c^{(6)}
_{\rm NLO}=0$ only holds when the LO matching condition is performed up to and
including terms of ${\cal O}(\varepsilon)$. In particular, $D$-dimensional
matching requires enlarging the operator basis to include (renormalized)
evanescent operators, which must be inserted in the one-loop diagrams of the
effective theory. Because of the IR divergences, the matrix elements of the
renormalized evanescent operators do not vanish in the $D\to 4$ limit.

A similar discussion concerning the matching procedure in
the presence of IR divergences can be found in ref.~\cite{Misiak:1999yg}. We mention
that the relevant evanescent operators, which enter the matching in our
calculation, are
\be
\begin{array}{ll}
E_1^q = \bar b \gamma_\mu \gamma_\alpha \gamma_{\nu\, L} b\; \bar q
\gamma^\nu \gamma^\alpha \gamma^\mu_L q -
4 \, O^q_1\,,&
E_3^q = \bar b \gamma_\mu \gamma_\alpha \gamma_{\nu\, L} t^a b\; \bar q
\gamma^\nu \gamma^\alpha \gamma^\mu_L t^a q -
4 \, O^q_3\,, \\
E_2^q = \bar b \gamma_\mu \gamma_\alpha \gamma_{\nu\, R} b\; \bar q
\gamma^\nu \gamma^\alpha \gamma^\mu_L q -
16 \, O^q_2\,,&
E_4^q = \bar b \gamma_\mu \gamma_\alpha \gamma_{\nu\, R} t^a b\; \bar q
\gamma^\nu \gamma^\alpha \gamma^\mu_L t^a q -
16 \, O^q_4\,.
\end{array}
\label{eq:ev}
\ee
The procedure for taking into account their contribution is further discussed
in the Appendix. We have also checked, on a proper subset of gauge-invariant
diagrams, that a calculation using the gluon mass as IR regulator gives the
same result as the one obtained with dimensional regularization by performing
the matching at ${\cal O}(\varepsilon)$, and including the insertion of the
evanescent operators.

We conclude this section by mentioning that our calculation of the Wilson
coefficients has passed the following checks:
\begin{itemize}
\item {\it gauge invariance}:
we verified that the coefficients are explicitly gauge-invariant. The same is
true for the full and the effective amplitudes separately;
\item {\it renormalization-scale dependence}:
the coefficient functions have the correct re\-nor\-ma\-li\-za\-tion-group
behaviour as predicted by the LO anomalous dimension matrix of the $\DB=0$
operators;
\item {\it IR divergences}: the coefficient functions are infrared finite. The
cancellation of IR divergences also takes place for the abelian combination of
diagrams in both the full and the effective amplitudes.
\end{itemize}

%%%%%%%%%%%%%%%%%%%%%%%%%%%%%%%%%%%%%%%%%%%%%%%%%%%%%%%%%%%%
\section{Spectator contributions at the NLO}
\label{sec:result}
%%%%%%%%%%%%%%%%%%%%%%%%%%%%%%%%%%%%%%%%%%%%%%%%%%%%%%%%%%%%
In this section we present the results for the coefficient functions of  the
dimension-six operators $O_k^q$ of eq.~(\ref{eq:opeff}) for $q=u,d,s$.
We collect the known LO expressions and then present our new results for the NLO
contributions.

The width operator $\widehat \Gamma$ in eq.~(\ref{eq:master}) depends
quadratically on  the coefficient functions $C_i$ of the $\DB=1$ effective
hamiltonian. Therefore, we find convenient to write the coefficient functions of
the dimension-six operators as
\be
c^q_k(\mu) = \sum_{i,j=1,2} C_i(\mu_1)\, C_j(\mu_1)\,
F^q_{k,ij}(\mu_1,\mu)\,,
\label{eq:c6db0}
\ee
where $\mu_1$ is the renormalization scale of the $\DB=1$ effective hamiltonian
and $\mu$ is the renormalization scale of the $\DB=0$ operators entering the
expansion of $\widehat \Gamma$~\footnote{~In the case $q=s$, at the NLO, there
are other contributions to eq.~(\ref{eq:c6db0}) coming from the insertion of
$\DB=1$ penguin operators ($i,j > 2$) which will be discussed at the end of the
section.}. The coefficients $F^q_{k,ij}$ depend on the renormalization scheme
and scale of both the $\DB=0$ and $\DB=1$ operators. The
dependence on $\mu_1$ and on the renormalization scheme of the $\DB=1$
operators actually cancels, order by order in perturbation theory, against the
corresponding dependence of the $\DB=1$ Wilson coefficients $C_i$. Therefore,
the coefficient functions $c^q_k$ only depend on the renormalization
scheme of the $\DB=0$ operators. A given scheme of the $\DB=1$
operators, however, must be chosen in order to present results for $F^q_{k,ij}$. In the
following we always consider $\DB=1$ operators renormalized in the $\msb$
scheme, defined in details in ref.~\cite{db2nlo1}. The corresponding
Wilson coefficients can be found in refs.~\cite{nlodb1a}-\cite{nlodb1c}.

Since we use dimensional regularization to regularize both UV and IR
divergences, the scales $\mu$ and
$\mu_1$ in eq.~(\ref{eq:c6db0}) play the r\^ole of both renormalization scale
and IR regulator. Dimensionally-regularized IR divergences produce additional
$\log\mu$ and $\log\mu_1$, beside those governed by the renormalization-group
equations. In the matching procedure, the same IR regulators in the full and
effective theories must be used, therefore we have to identify $\mu$ and
$\mu_1$. This choice gets rid of the spurious $\log(\mu/\mu_1)$ so that the NLO
coefficient functions have the correct UV behaviour predicted by the LO
anomalous dimensions of the $\DB=0$ operators.

\begin{figure} [t]
\begin{center}
%\vspace*{-.55cm}
\epsfxsize=14cm
\epsfbox{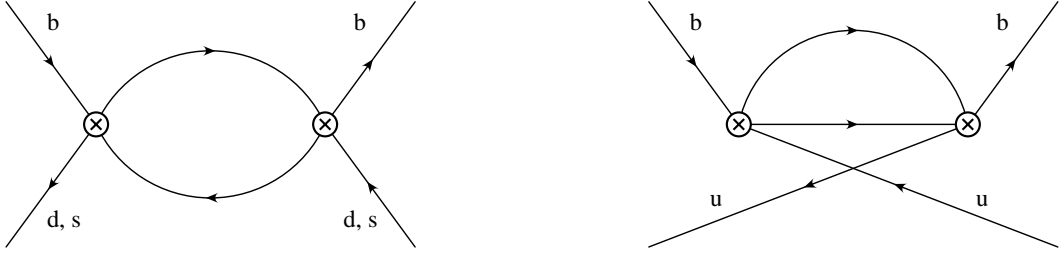}
%\vspace*{-1.3cm}
\end{center}
\caption{\it Feynman diagrams entering at the LO in the calculation of the
coefficient functions $c^q_k$ for $q=u,d,s$.}
\label{fig:lospec}
\end{figure}

For the coefficients $F^q_{k,ij}$ we write the expansion
\be
F^q_{k,ij}(\mu,\mu) = A^{q}_{k,ij} + \frac{\as}{4\pi} B^{q}_{k,ij}(\mu)\,.
\ee
Since, by definition, the coefficients $F^q_{k,ij}$ are symmetric in the indices
$i$ and $j$, we will only present results for $i \le j$. The leading order
coefficients $A^{d}_{k,ij}$ have been computed with a non-vanishing charm quark
mass by Neubert and Sachrajda~\cite{NS}. The relevant one-loop diagram is shown
in fig.~\ref{fig:lospec} (left). We have repeated the calculation
and found results in agreement with ref.~\cite{NS}. In the basis of eq.~(\ref{eq:opeff}), the
coefficients $A^{q}_{k,ij}$, for the case $q=d$, read
\be
\begin{array}{ll}
A^d_{1,11} = -\dfrac{1}{3}\left( 1 + {\dfrac{z}{2}} \right) \,
{{\left( 1 - z \right) }^2}\,, &
A^d_{1,12} = -\dfrac{1}{9}\left( 1 + {\dfrac{z}{2}} \right) \,
{{\left( 1 - z \right) }^2} \,, \\
 A^d_{1,22} = -\dfrac{1}{3} \left( 1 + {\dfrac{z}{2}} \right) \,
{{\left( 1 - z \right) }^2}\,, &
A^d_{2,11} = -\dfrac{1}{6} {{\left( 1 - z \right) }^2}\,
\left( 1 + 2\,z \right) \,,  \\
 A^d_{2,12} = -\dfrac{1}{18}{{\left( 1 - z \right) }^2}\,
\left( 1 + 2\,z \right) \,,\quad  &
 A^d_{2,22} = -\dfrac{1}{6} {{\left( 1 - z \right) }^2}\,
\left( 1 + 2\,z \right) \,, \\
A^d_{3,11} = -2\left( 1 + {\dfrac{z}{2}} \right) \,{{\left( 1 - z \right)
}^2}\,, &
A^d_{3,12} = -\dfrac{2}{3} \left( 1 + {\dfrac{z}{2}} \right)
\,{{\left( 1 - z \right) }^2}\,, \\
 A^d_{3,22} = 0\,, &
A^d_{4,11} = -{{\left( 1 - z \right) }^2}\,\left( 1 + 2\,z \right)\,, \\
 A^d_{4,12} = -\dfrac{1}{3} {{\left( 1 - z \right) }^2}\,
\left( 1 + 2\,z \right) \,,  &
 A^d_{4,22} = 0\, ,
\end{array}
\ee
where $z=m_c^2/m_b^2$.

At the LO, the only difference between the case $q=s$ and $q=d$ is the massive
charm quark running in the loop. The coefficients $A^s_{k,ij}$ are given by
\be
\begin{array}{ll}
A^s_{1,11} = -\dfrac{1}{3}\sqrt{1-4z} \,
{{\left( 1 - z \right) }}\,, &
A^s_{1,12} = -\dfrac{1}{9}\sqrt{1-4z} \,
{{\left( 1 - z \right) }} \,, \\
 A^s_{1,22} = -\dfrac{1}{3}\sqrt{1-4z} \,
{\left( 1 - z \right) }\,, &
A^s_{2,11} = -\dfrac{1}{6}\sqrt{1-4z}\,
\left( 1 + 2\,z \right) \,,  \\
 A^s_{2,12} = -\dfrac{1}{18}\sqrt{1-4z}\,
\left( 1 + 2\,z \right) \,,\quad  &
 A^s_{2,22} = -\dfrac{1}{6}\sqrt{1-4z}\,
\left( 1 + 2\,z \right) \,, \\
A^s_{3,11} = -2\sqrt{1-4z} \,{{\left( 1 - z \right)
}}\,, &
A^s_{3,12} = -\dfrac{2}{3}\sqrt{1-4z}
\,{{\left( 1 - z \right) }}\,, \\
 A^s_{3,22} = 0\,, &
A^s_{4,11} = -\sqrt{1-4z}\,\left( 1 + 2\,z \right)\,, \\
 A^s_{4,12} = -\dfrac{1}{3}\sqrt{1-4z}\,
\left( 1 + 2\,z \right) \,,  &
 A^s_{4,22} = 0\,.
\end{array}
\ee

%________________________________________________________________
%________________________________________________________________

For $q=u$ instead, the relevant one-loop diagram is shown in
fig.~\ref{fig:lospec} (right). It gives
\be
\begin{array}{lll}
A^u_{1,11} = {{\left( 1 - z \right) }^2}\,, \quad &
 A^u_{1,12} = \dfrac{1}{3}\left( 1 - z \right)^2 \,, \quad  &
 A^u_{1,22} = {{\left( 1 - z \right) }^2}\,, \\
 A^u_{2,11} = 0\,,  &
 A^u_{2,12} =0 \,,  &
 A^u_{2,22} = 0\,, \\
 A^u_{3,11} = 0\,, &
 A^u_{3,12} = 2\,{{\left( 1 - z \right) }^2}\,,  &
 A^u_{3,22} = 0\,, \\
 A^u_{4,11} = 0\,, &
 A^u_{4,12} = 0 \,,  &
 A^u_{4,22} = 0\,.
\end{array}
\ee

The NLO coefficients $B^{d}_{k,ij}$ are obtained by computing the diagrams shown
in figs.~\ref{fig:fddb1} and~\ref{fig:diaeff}. The details of the matching
procedure have been discussed in the previous section. We just remind here that
we have neglected corrections of ${\cal O}(m_c^2/m_b^2)$ in the NLO calculation.

At the NLO, the coefficients depend on the renormalization scheme chosen for
the HQET operators in eq.~(\ref{eq:opeff}). We have computed these coefficients
in the $\msb$ scheme defined in ref.~\cite{reyes}. In this scheme, for the
coefficients $B^{d}_{k,ij}(\mu)$ we obtain
\be
\begin{array}{lll}
B^d_{1,11} = -{\dfrac{230}{81}}
\,, &
B^d_{1,12} = {\dfrac{778}{243}} + {\dfrac{32\,{L}}{9}}
\,, &
B^d_{1,22} = -{\dfrac{314}{81}} + {\dfrac{16\,{{\pi }^2}}{27}} \,,\\

B^d_{2,11} = -{\dfrac{100}{81}}
\,, &
B^d_{2,12} = {\dfrac{404}{243}} + {\dfrac{16\,{L}}{9}}
\,, &
B^d_{2,22} = -{\dfrac{106}{81}} + {\dfrac{8\,{{\pi }^2}}{27}}  \,, \\

B^d_{3,11} = {\dfrac{925}{54}} + 18\,{L}
\,, &
B^d_{3,12} = {\dfrac{2689}{162}} + {\dfrac{46\,{L}}{3}}
\,, &
B^d_{3,22} ={\dfrac{91}{9}} - {\dfrac{4\,{{\pi }^2}}{9}} + 8\,{L} \,, \\

B^d_{4,11} = {\dfrac{455}{54}} + 9\,{L}
\,, \quad &
B^d_{4,12} = {\dfrac{1337}{162}} + {\dfrac{23\,{L}}{3}}
\,, \quad &
B^d_{4,22} ={\dfrac{85}{18}} - {\dfrac{2\,{{\pi }^2}}{9}} + 4\,{L} \,,
\end{array}
\ee
where $L = \log \left(\mu/m_b\right)$.
In the case $q=u$, the relevant Feynman diagrams involve gluon corrections
to the LO diagram shown on the right side of fig.~\ref{fig:lospec}. We obtain
for the coefficients
\be
\begin{array}{ll}
B^u_{1,11} = {\dfrac{188}{27}} - {\dfrac{16\,{{\pi }^2}}{9}}
\,, &
B^u_{1,12} = -{\dfrac{1312}{81}} - {\dfrac{32\,{{\pi }^2}}{27}} -
{\dfrac{32\,{L}}{3}}
\, , \\
B^u_{1,22} = {\dfrac{188}{27}} - {\dfrac{16\,{{\pi }^2}}{9}}
\, , &
B^u_{2,11} = {\dfrac{16}{27}}
\, , \\
B^u_{2,12} = -{\dfrac{32}{81}}
\, , &
B^u_{2,22} = {\dfrac{16}{27}}
\, , \\
B^u_{3,11} = -{\dfrac{125}{3}} - {\dfrac{4\,{{\pi }^2}}{3}} - 24\,{L}
\, , \quad &
B^u_{3,12} = -{\dfrac{374}{27}} - {\dfrac{28\,{{\pi }^2}}{9}} - 10\,{L}
\, , \\
B^u_{3,22} = -{\dfrac{125}{3}} - {\dfrac{4\,{{\pi }^2}}{3}} - 24\,{L}
\, , &
B^u_{4,11} = -{\dfrac{4}{3}}
\, , \\
B^u_{4,12} = -{\dfrac{100}{27}}
\, , &
B^u_{4,22} = -{\dfrac{4}{3}}
\,.
\end{array}
\ee

\begin{table} [t]
\begin{center}
\begin{tabular}{||l||c|c||c|c||c|c||}
\cline{2-7}
\multicolumn{1}{c||}{} & \multicolumn{2}{c||}{$q=d$} &
\multicolumn{2}{c||}{$q=u$} & \multicolumn{2}{c||}{$q=s$} \\
\cline{2-7}
\multicolumn{1}{c||}{} & LO & NLO &  LO & NLO & LO & NLO \\
\hline
$c^q_1$ &  $-0.32$& $-0.30$ & $ 0.92$ & $ 0.90$& $-0.27$ & $-0.25$ \\
$c^q_2$ &  $-0.18$& $-0.16$ & $ 0.00$ & $0.02$& $-0.17$ & $-0.15$ \\
$c^q_3$ &   $0.20$& $ 0.16$ & $-0.83$ & $-1.42$& $ 0.17$ & $0.13$ \\
$c^q_4$ &   $0.11$& $ 0.08$ & $ 0.00$ & $0.00$ & $ 0.11$ & $0.08$ \\
\hline \hline
$\widetilde c^{\,q}_1$ & $-0.02 $ & $-0.03 $ & $ -0.06$ & $-0.33 $ & $ -0.01$ & $-0.03 $ \\
$\widetilde c^{\,q}_2$ & $ 0.02$  & $0.03$    & $0.00 $  & $-0.01 $ & $0.02 $   & $0.03 $ \\
$\widetilde c^{\,q}_3$ & $-0.70 $ & $-0.65 $ & $2.11 $  & $2.27 $   & $-0.60 $  & $-0.54 $ \\
$\widetilde c^{\,q}_4$ & $0.79 $  & $0.68$    & $0.00 $  & $-0.06 $   & $0.77 $  & $0.65 $ \\
\hline
\end{tabular}
\end{center}
\caption{\it Wilson coefficients $c^q_k$ computed at the LO and NLO
($\overline{\rm MS}$ scheme). As reference values, we use $\mu=m_b=4.8$ GeV and
$m_c=1.4$ GeV.  The coefficients $\widetilde c^q_k$ in the operator basis of
eq.~(\ref{eq:newba}) are also shown.}
\label{tab:coeff}
\end{table}

Finally, we discuss the extension of eq.~(\ref{eq:c6db0}) to the case $q=s$.
With respect to $c^d_k(\mu)$, the coefficient functions $c^s_k(\mu)$
receive, at the NLO, additional contributions coming from the penguin
contraction of the current-current operators (diagram $D_{13}$ of
fig.~\ref{fig:fddb1}) and from the insertion of penguin and chromomagnetic
operators (diagram $D_{14}$ of fig.~\ref{fig:fddb1}). Since the Wilson
coefficients $C_3$--$C_6$ are small, contributions with a double
insertion of penguin operators can be safely neglected. As
suggested in \cite{Beneke:1999sy}, a consistent way for implementing this
approximation is to consider the coefficients $C_3$--$C_6$ as
formally of ${\cal O}(\alpha_s)$. Within this approximation, only
single insertions of penguin operators need to be
considered at the NLO. Therefore we can write
\renewcommand{\arraystretch}{1.0}
\bea
c^s_k(\mu) &=& c^d_k(\mu) + \frac{\as}{4\pi} \,
C_2(\mu)^2 \, P_{k,22}(\mu) + 2\frac{\as}{4\pi}C_2(\mu)C_{8G}(\mu) P_{k,28}(\mu)
+\nn\\
&&2\dsum_{i=1,2} \dsum_{r=3,4}
C_i(\mu) \, C_r(\mu) P_{k,ir}(\mu)\,,
\eea
and we obtain
\renewcommand{\arraystretch}{2.0}
\be
\begin{array}{ll}
P_{1,22} = \dfrac{32}{243}  + \dfrac{32\,{L}}{81} \,, \quad &
P_{2,22} = \dfrac{16}{243}  + \dfrac{16\,{L}}{81} \, , \\
P_{3,22} = -\dfrac{8}{81} - \dfrac{8\,{L}}{27} \, , &
P_{4,22} = -\dfrac{4}{81} - \dfrac{4\,{L}}{27} \, ,
\end{array}
\ee
\be
\begin{array}{llll}
P_{1,13} = -\dfrac{1}{3} \,, & P_{2,13} = -\dfrac{1}{6} \,, &
P_{3,13} = -2 \,, & P_{4,13} = -1\,, \\
P_{1,14} = -\dfrac{1}{9}\,, & P_{2,14} = -\dfrac{1}{18} \,, &
P_{3,14} = -\dfrac{2}{3}\,, & P_{4,14} = -\dfrac{1}{3}\,, \\
P_{1,23} = -\dfrac{1}{9}\,, & P_{2,23} = -\dfrac{1}{18} \,, &
P_{3,23} = -\dfrac{2}{3}\,, & P_{4,23} = -\dfrac{1}{3}\,, \\
P_{1,24} = -\dfrac{1}{3}\,, & P_{2,24} = -\dfrac{1}{6} \,, &
P_{3,24} = 0 \,, & P_{4,24} = 0 \,, \\
P_{1,28} = -\dfrac{8}{27} \,,\quad & P_{2,28} = -\dfrac{4}{27} \,,\quad &
P_{3,28} = \dfrac{2}{9} \,,\quad & P_{4,28} = \dfrac{1}{9}  \,.
\end{array}
\ee
The coefficients $P_{k,28}$ are computed using the convention in
which the chromomagnetic coefficient $C_8$ has a positive sign. The NLO
contribution of penguin and chromomagnetic operators to beauty hadron
lifetimes has been also computed in ref.~\cite{keum}. We verified that our
results are in agreement with them.

For convenience, we present in table \ref{tab:coeff} the numerical values of the
coefficients $c_k^q(\mu)$, in the cases $q=u,d,s$, at $\mu=m_b=4.8$ GeV,
both at LO and NLO.  We also give the coefficients  $\widetilde c_k^q(\mu)$
defined in the operator basis of ref.~\cite{NS}.  For details on the conversion
between the two basis, see sect.~\ref{sec:ME}

%%%%%%%%%%%%%%%%%%%%%%%%%%%%%%%%%%%%%%%%%%%%%%%%%%%%%%%%%%%%%%%%%%%%%%
\section{Matching of $\DB=0$ non-singlet operators between QCD and HQET}
\label{sec:match}
%%%%%%%%%%%%%%%%%%%%%%%%%%%%%%%%%%%%%%%%%%%%%%%%%%%%%%%%%%%%
Theoretical predictions of beauty hadron lifetimes, based on the OPE, are
obtained by combining the perturbative determination of the Wilson coefficients
with the non-perturbative evaluation of the hadronic matrix elements of the HQET
operators. In the next section, we will perform a phenomenological
analysis, by using the NLO results for the Wilson coefficients and the lattice
determinations of the relevant hadronic matrix elements. In some cases
the lattice results are provided in terms of matrix elements of QCD
operators. In order to use these determinations, consistently at the NLO, it is
necessary to perform an additional matching between the QCD and the HQET
operators. For the dimension-six $\DB=0$ operators of interest in this paper,
this matching has not been computed so far. For this reason, we present in this
section the results of an ${\cal O}(\as)$ calculation of the coefficient
functions relating the QCD four-fermion $\DB=0$ operators to their counterparts
in HQET.

We limit ourselves to the case of flavour non-singlet operators, which
are the only ones for which lattice results are available so far. An important
example is the difference $O^u_k-O^d_k$ of the operators defined in
eq.~(\ref{eq:opeff}) which determines, as we will see below, the lifetime ratio
of charged to neutral $B$ mesons. In the limit of exact flavour symmetry, these
operators do not mix with lower dimensional operators. For this reason, their
matching between QCD and HQET does not require the calculation of penguin
contractions and only involves current-current operators of dimension six.

The matching equation between the operators in QCD and HQET can be written in
the form
\be
{\vec O}_{\rm QCD}(\mu) = \widehat C(\mu,\mu',m_b) \, {\vec O}_{\rm HQET}(\mu')
\, .
\ee
The scale dependence of the coefficient functions $\widehat C(\mu,\mu',m_b)$ is
governed by the renormalization group equation of the QCD and HQET operators
respectively. Thus one finds
\be
\widehat C(\mu,\mu',m_b) = \left( W(\mu,m_b)^T\right)^{-1} \widehat C(m_b) \,
\widetilde W(\mu',m_b)^T
\label{eq:cmu}
\ee
where the matrix $W(\mu,m_b)$ in QCD is given, at the NLO, by
\be
\label{eq:W}
W(\mu,m_b) = \left(1+ \frac{\as(\mu)}{4\pi} J(\mu)\right)
\left( \frac{\as(\mu)}{\as(m_b)} \right)^{-\gamma_0^T/2 \beta_0}
\left(1- \frac{\as(m_b)}{4\pi} J(m_b)\right)
\ee
and $\widetilde W(\mu,m_b)$ is its analogous in the HQET. For convenience, we
present here the expressions of the leading anomalous dimension matrices,
$\gamma_0$ and $\widetilde\gamma_0$ in QCD and HQET. In the basis of
eq.~(\ref{eq:opeff}) they reads~\cite{Bagger:1997gg,Chernyak:1995cx}%
\footnote{Note that the expression of the LO
anomalous dimension matrix in HQET given in eq.(A.1) of ref.\cite{NS} contains
a misprint: the matrix element $(\widetilde\gamma_0)_{12}$ should be read as
-1 instead of 1.}
\renewcommand{\arraystretch}{1.2}
\be
\gamma_0=
\left(\begin{array}{cccc}
    0 & 0 & 12 & 0 \cr
    0 & 0 & 0 & -12 \cr
    8/3 & 0 & -4 & 0 \cr
    0 & -8/3 & 0 & -14 \cr
\end{array}\right) \, , \qquad
\widetilde\gamma_0=
\left(\begin{array}{cccc}
   0 & 0 & 0 & 0 \cr
   0 & 0 & 0 & 0 \cr
   0 & 0 &-9 & 0 \cr
   0 & 0 & 0 & -9 \cr
\end{array}\right) \, .
\ee
\renewcommand{\arraystretch}{2.0}
The matrix $J$, in eq.~(\ref{eq:W}), is expressed in terms of the two-loop
anomalous dimension, and it is defined for instance in ref.~\cite{db2nlo1}. In
the case of the HQET, the NLO anomalous dimension is still unknown.

The matrix $\widehat C(m_b)$ is the coefficient function at the matching scales
$\mu=\mu'=m_b$. At the NLO, it can be written in the form:
\be
\label{eq:cmatch}
\widehat C(m_b) = 1 + \frac{\as(m_b)}{4\pi} \, \widehat C_1 \, .
\ee
We have computed $\widehat C_1$ considering two different renormalization
schemes for the QCD operators, namely $\msb$ and Landau $\ri$ (see
ref.~\cite{db2nlo1} for a detailed definitions of these schemes). The HQET
operators are renormalized in the $\msb$ scheme of ref.~\cite{reyes}. By
denoting the results with $\widehat C_1^{\overline{\rm{MS}}}$ and $\widehat
C_1^{\rm RI}$ respectively, in the operator basis of eq.~(\ref{eq:opeff}), we
obtain:
\renewcommand{\arraystretch}{1.2}
\be
\label{eq:c1ms}
\widehat C_1^{\overline{\rm{MS}}} =
\left(\begin{array}{cccc}
  - 4/3 & 4/3	& -18	& -2  \cr
    4/3 & - 4/3 & 2	& 6 \cr
     -4 & -4/9  & -35/6 & -5/2 \cr
    4/9 & 4/3	& - 5/6 & 25/6 \cr
\end{array}\right)
\ee
and
\be
\widehat C_1^{\rm RI} =
\left(\begin{array}{cccc}
    - 4/3   & 4/3 & -4 - 24\,{\log2} & -2 \cr
      4/3 & - 4/3   & 2 & 2 - 4\,{\log2} \cr
      - 8/9   - 16/3\,{\log2} & - 4/9   & - 21/2 + 8\,{\log2} & - 5/2 \cr
      4/9 & 4/9 - 8/9\,{\log2} & - 5/6   & - 13/2 + 4/3\,{\log2} \cr
\end{array}\right) \, .
\ee
\renewcommand{\arraystretch}{2.0}

Note that the difference $C_1^{\overline{\rm{MS}}} - C_1^{\rm RI}$ is the one
loop matrix which provides the connection between the $\overline{\rm{MS}}$ and
$\rm RI$ schemes in QCD for the $\DB=0$ operators. We have checked that this
matrix agrees with the matrix denoted as $\hat r_{\overline{\rm{MS}}}$ in
ref.~\cite{db2nlo1}, once the proper change of basis is performed.

In the next section, the matrix $C_1^{\overline{\rm{MS}}}$ of
eq.~(\ref{eq:c1ms}) will be used to convert the lattice results for the matrix
elements of the QCD operators into their HQET counterparts.

%%%%%%%%%%%%%%%%%%%%%%%%%%%%%%%%%%%%%%%%%%%%%%%%%%%%%%%%%%%%
\section{Phenomenological Discussion}
\label{sec:phenom}
%%%%%%%%%%%%%%%%%%%%%%%%%%%%%%%%%%%%%%%%%%%%%%%%%%%%%%%%%%%%

In this section we perform a phenomenological analysis of the lifetime ratios
of beauty hadrons, by using the NLO expressions of the Wilson coefficients and
the lattice determinations of the relevant hadronic matrix
elements~\cite{DiPierro98}-\cite{APE}.

The starting point of this analysis is eq.~(\ref{eq:ratio}), which expresses the
ratio of inclusive widths of beauty hadrons, up to and including $1/m_b^3$
spectator effects. The combinations of hadronic parameters entering this
formula at order $1/m_b^2$ can be evaluated from the heavy hadron
spectroscopy~\cite{mupig}. The numerical contributions of these terms to the
lifetime ratios are rather small, and one obtains the estimate
\be
\label{eq:del}
\dfrac{\tau(B^+)}{\tau(B_d)}= 1.00 -\Delta^{B^+}_{\rm\scriptstyle spec}\,,\quad
\dfrac{\tau(B_s)}{\tau(B_d)}= 1.00 -\Delta^{B_s}_{\rm\scriptstyle spec}\,,\quad
\dfrac{\tau(\Lambda_b)}{\tau(B_d)}= 0.98(1)-
\Delta^{\Lambda}_{\rm\scriptstyle spec} \,,
\ee
which can be compared with the experimental results in eq.~(\ref{eq:rexp}). In
the previous expressions the $\Delta$s represent the $1/m_b^3$ contributions of
hard spectator effects
\be
\label{eq:delta}
\Delta^{H_b}= \dfrac{96\pi^2}{m_b^3\, c^{(3)}}\dsum\limits_{k}
c^{(6)}_k \left(\frac{\langle O^{(6)}_k\rangle_{H_b}}{2M_{H_b}}-\frac{\langle
O_k^{(6)}\rangle_{B_d}}{2M_{B_d}}\right) \, .
\ee
These are the quantities which we are interested in. They are expressed in
terms of coefficient functions and matrix elements of dimension-six operators.
In the phenomenological analysis, we use the non-perturbative determination
of the relevant matrix elements from lattice QCD calculation. For recent estimates based
on QCD sum rules, see refs.~\cite{Colangelo:1996ta}--\cite{Huang:1999xj}.

\subsection{Matrix Elements}
\label{sec:ME}
The basis of four-fermion operators usually considered in the
literature to study the lifetime ratios of beauty hadrons is given by~\cite{NS}
\be
\begin{array}{ll}
{\cal O}^q_1= (\bar b_i q_i)_{V-A} \, (\bar q_j b_j)_{V-A}\,, &
{\cal O}^q_2= (\bar b_i q_i)_{S-P} \, (\bar q_j b_j)_{S+P}\,, \\
{\cal O}^q_3= (\bar b_i t^{a}_{ij} q_j)_{V-A} \,
(\bar q_k t^{a}_{kl} b_l)_{V-A}\,, &
{\cal O}^q_4= (\bar b_i t^{a}_{ij} q_j)_{S-P} \,
(\bar q_k t^{a}_{kl} b_l)_{S+P}\, ,
\end{array}
\label{eq:newba}
\ee
where $q=u,d,s,c$ and $(\bar q q)_{S\pm P}=\bar q (1\pm \gamma^5)q$. In this
basis, which differs from the one in eq.~(\ref{eq:opeff}) by a Fierz
rearrangement, the matrix elements of the operators ${\cal O}^q_3$ and
${\cal O}^q_4$, between external $B$-mesons states, vanish in the vacuum
saturation approximation (VSA). By following the convention adopted in the
literature, we will work in this section in the basis of
eq.~(\ref{eq:newba}), to which we add the penguin operator of
eq.~(\ref{eq:openg}). The Wilson coefficients associated to the new basis will
be indicated by the symbol $\widetilde c_k^{\, q}$. They are related to those
defined in the previous sections by a linear transformation
\renewcommand{\arraystretch}{1.2}
\be
\label{eq:ctilde}
\widetilde c_k^{\, q} = M_{kj} c_j^q \, , \qquad
M=\left(\begin{array}{cccc}
    1/3 & 0 & 4/9 & 0 \cr
    0 & -2/3 & 0 & -8/9 \cr
    2 & 0 & -1/3 & 0 \cr
    0 & -4 & 0 & 2/3 \cr
\end{array}\right) \, .
\ee
\renewcommand{\arraystretch}{2.0}
The values of $ \widetilde c_k^{\, q}$ at $\mu=m_b$ are given in table
\ref{tab:coeff}.

\begin{figure} [t]
\begin{center}
%\vspace*{-.55cm}
\epsfxsize=14cm
\epsfbox{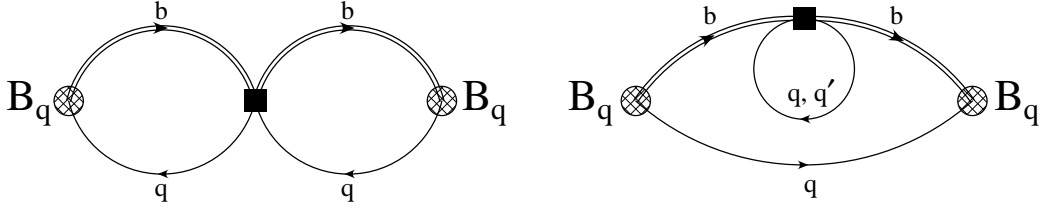}
%\vspace*{-1.3cm}
\end{center}
\caption{\it Feynman diagrams entering the non-perturbative evaluation of the
matrix elements of valence operators between external $B$-meson states. In the
case of non-valence operators only the second diagram appears.}
\label{fig:aie}
\end{figure}

In parametrizing the matrix elements of the current-current operators, it is
useful to distinguish two cases, depending on whether or not the light quark
$q$ of the operator enters as a valence quark in the external hadronic state.
From a diagrammatic point of view, this difference gives rise to a different
number of Wick contractions. In the case of external $B$-meson states, for
instance, the matrix elements of the valence operators are computed by
evaluating non-perturbatively the two Feynman diagrams shown in
fig.~\ref{fig:aie}, while for non-valence operators only the second diagram
appears.
%____________________________________________________________________________
%___________________________________________________________________________
We find it convenient to introduce different $B$-parameters for the valence and
non-valence contributions. Thus, for the $B$-meson matrix elements of the
non-valence operator we define
\be
\dfrac{\langle B_{q}\vert {\cal O}^{q'}_k\vert B_{q}\rangle}{2M_{B_q}}=
\dfrac{f_{B_q}^2M_{B_q}}{2} \, \delta_k^{\,q'q} \quad {\rm for} \,\, q \neq q'
\,,
\label{eq:bpar1}
\ee
and, for the valence operators ($q=q'$), we write
\be
\begin{array}{ll}
\dfrac{\langle B_q\vert {\cal O}^q_1 \vert B_q\rangle}{2M_{B_q}}=
\dfrac{f_{B_q}^2M_{B_q}}{2} \, \left( B_1^{\, q} +
\delta_1^{\, qq}\right) \, , \quad
\dfrac{\langle B_q\vert {\cal O}^q_3 \vert B_q\rangle}{2M_{B_q}}=
\dfrac{f_{B_q}^2M_{B_q}}{2} \, \left( \ep_1^{\, q} +
\delta_3^{\, qq}\right) \, , \\
\dfrac{\langle B_q\vert {\cal O}^q_2 \vert B_q\rangle}{2M_{B_q}}=
\dfrac{f_{B_q}^2M_{B_q}}{2} \, \left( B_2^{\, q} +
\delta_2^{\, qq}\right) \, , \quad
\dfrac{\langle B_q\vert {\cal O}^q_4 \vert B_q\rangle}{2M_{B_q}}=
\dfrac{f_{B_q}^2M_{B_q}}{2} \, \left( \ep_2^{\, q} +
\delta_4^{\, qq}\right) \,.
\end{array}
\label{eq:bpar2}
\ee
In eq.~(\ref{eq:bpar2}), the $\delta_k^{\, qq}$ are defined as the parameters
$\delta_k^{\, qq'}$ of eq.~(\ref{eq:bpar1}) in the limit of degenerate quark
masses ($m_q=m_{q'}$). In the VSA, $B_1^{\, q}=B_2^{\, q}=1$ while the $\ep$
parameters and all the $\delta$s vanish. It should be clear that the $B_k$,
$\ep_k$ and $\delta_k$ have a direct interpretation in terms of Feynman
diagrams, and are represented by the first ($B_k$, $\ep_k$) and second
($\delta_k$) contraction of fig.~\ref{fig:aie} respectively. Each of these
parameters is a well defined and gauge invariant quantity.
In the exact $SU(2)$ limit, for instance, the parameters $B_k^{\, d}$ express
the matrix elements of the non-singlet operator $O^u_k-O^d_k$ between external
$B$-meson states.

Our definition of $B_1$, $B_2$, $\ep_1$ and $\ep_2$ differs from the one
considered in the literature for the presence of the $\delta$s in
eq.~(\ref{eq:bpar2}). The reason why we have distinguished between valence and
non valence contributions, in these definitions, is that lattice QCD
calculations~\cite{DiPierro98}-\cite{APE} have not computed, so far, the full
matrix elements, but only the valence contributions represented the first
diagram of fig.~\ref{fig:aie}. Thus, the $B_k$ and $\ep_k$ parameters, defined
as in eq.~(\ref{eq:bpar2}), denote the quantities actually determined, at
present, by lattice calculations. In principle, a non-perturbative calculation of the
$\delta$ parameters on the lattice is possible. However, it requires dealing
with the difficult problem of power-divergence subtractions which has prevented
so far the calculation of the corresponding diagrams.

To complete the definitions of the $B$-parameters for the $B$-mesons, we
introduce a parameter for the matrix element of the penguin operator
\be
\dfrac{\langle B_q\vert O_P\vert B_q\rangle}{2M_{B_q}}=
\dfrac{f_{B}^2M_{B}}{2} \, P^{\, q} \,.
\label{eq:bpar3}
\ee

We now define the $B$-parameters for the $\Lambda_b$ baryon. Up to $1/m_b$
corrections, the matrix elements of the operators ${\cal O}^q_2$ and ${\cal
O}^q_4$, between external $\Lambda_b$ states, can be related to the matrix
elements of the operators ${\cal O}^q_1$ and ${\cal O}^q_3$~\cite{NS}
\be
\langle \Lambda_b\vert {\cal O}^q_1 \vert \Lambda_b\rangle = -2 \,
\langle \Lambda_b\vert {\cal O}^q_2 \vert \Lambda_b\rangle \, , \quad
\langle \Lambda_b\vert {\cal O}^q_3 \vert \Lambda_b\rangle = -2 \,
\langle \Lambda_b\vert {\cal O}^q_4 \vert \Lambda_b\rangle \, .
\ee
For the independent matrix elements, assuming $SU(2)$ symmetry, we define
\bea
\label{eq:bpar4}
&& \dfrac{\langle \Lambda_b\vert {\cal O}^q_1 \vert \Lambda_b\rangle}
{2M_{\Lambda_b}}= \dfrac{f_{B}^2M_{B}}{2} \, \left( L_1 +
\delta_1^{\,\Lambda q}\right) \quad {\rm for} \,\, q=u,d \, ,\nonumber \\
&& \dfrac{\langle \Lambda_b\vert {\cal O}^q_3 \vert \Lambda_b\rangle}
{2M_{\Lambda_b}}= \dfrac{f_{B}^2M_{B}}{2} \, \left( L_2 +
\delta_2^{\,\Lambda q}\right) \quad {\rm for} \,\, q=u,d  \, , \nonumber \\
&& \dfrac{\langle \Lambda_b\vert {\cal O}^q_1\vert \Lambda_b\rangle}
{2M_{\Lambda_b}}= \dfrac{f_{B}^2M_{B}}{2} \, \delta_1^{\, \Lambda q}
\quad {\rm for}\,\, q=s,c   \, ,\\
&& \dfrac{\langle \Lambda_b\vert {\cal O}^q_3\vert \Lambda_b\rangle}
{2M_{\Lambda_b}}= \dfrac{f_{B}^2M_{B}}{2} \, \delta_2^{\, \Lambda q}
\quad {\rm for}\,\, q=s,c   \, ,\nonumber \\
&& \dfrac{\langle \Lambda_b\vert O_P\vert \Lambda_b\rangle}{2M_{\Lambda_b}}=
\dfrac{f_{B}^2M_{B}}{2} \, P^{\, \Lambda} \nonumber \,.
\eea
In analogy with the $B$-meson case, the parameters $L_1$ and $L_2$ represent
the valence contributions computed by current lattice
calculations~\cite{DiPierro:1999tb,DiPierroproc}.

At present, two independent lattice calculations of the $B$-parameters in the
$B$-meson sector have been performed, both in the quenched approximation. In
the first study~\cite{DiPierro98,DiPierroproc}, the parameters $B_1^d$,
$B_2^d$, $\ep_1^d$ and $\ep_2^d$ have been computed by simulating on the
lattice the HQET. The second calculation~\cite{APE}, instead, has been
performed by using QCD, and the results have been obtained in this case for
both $B_{u,d}$ and $B_s$ mesons. The calculation of ref.\cite{APE} is also
accurate at the NLO, since the $B$-parameters have been evolved, from the
lattice scale up to $m_b$, by using the two-loop anomalous dimension of the
four-fermion $\DB=0$ operators, which is known in QCD~\cite{db2nlo1,db2nlo2},
but not in the HQET.
For these reasons, we will use the results of ref.~\cite{APE} in
our phenomenological analysis. The parameters $B_1^q$, $B_2^q$, $\ep_1^q$ and
$\ep_2^q$, computed in QCD, have been matched from QCD to HQET, at the NLO,
using the coefficients presented in the previous section,
eqs.~(\ref{eq:cmatch})-(\ref{eq:c1ms}). After applying these coefficients to
the QCD lattice results of ref.~\cite{APE}, we obtain the estimates of the
HQET $B$-parameters collected in table~\ref{tab:inputs}.
\begin{table} [t]
\renewcommand{\arraystretch}{1.6}
\begin{center}
\begin{tabular}{|cc|}
\hline
$ B_1^d$ = $1.2 \pm 0.2$ & $ B_1^s$ = $1.0 \pm 0.1$  \\
$ B_2^d$ = $0.7 \pm 0.1$ & $ B_2^s$ = $0.7 \pm 0.1$ \\
$ \ep_1^d$ = $0.03 \pm 0.02$ & $ \ep_1^s$ = $0.03 \pm 0.01$ \\
$ \ep_2^d$ = $0.04 \pm 0.01$ & $ \ep_2^s$ = $0.03 \pm 0.01$ \\
\hline
$ L_1$ = $-0.2 \pm 0.1$ & $ L_2$ = $0.2 \pm 0.1$ \\
\hline
$m_b$ = $4.8\pm 0.1$ GeV & $m_b-m_c$ = $3.40\pm 0.06$ GeV \\
$f_B$ = $200\pm 25$ MeV & $f_{B_s}/f_B$ = $1.16\pm 0.04$ \\
\hline
\end{tabular}
\end{center}
\caption{\it Central values and standard deviations of the input parameters used
in the numerical analysis.}
\label{tab:inputs}
\end{table}

For comparison, we also present here the values of $B$-parameters obtained in
ref.~\cite{DiPierro98,DiPierroproc}
from the lattice simulation in the HQET. In this case, a different definition
of the $\overline{\rm{MS}}$ scheme  has been adopted. The NLO connection
between this scheme and the one of ref.~\cite{reyes}, chosen in this paper, is
given by~\cite{reyes2}
\renewcommand{\arraystretch}{1.2}
\be
\label{eq:schemi}
\left(\begin{array}{c} B_2^q \cr \ep_2^q \end{array}\right) =
\left[ 1 + \frac{\as}{4\pi}
\left(\begin{array}{cc} 0 & -3/2 \cr -1/3 & -7/4 \end{array}\right)
\right]
\left(\begin{array}{c} B_2^q \cr \ep_2^q \end{array}\right)_{\rm SDP}
\ee
\renewcommand{\arraystretch}{2.0}
where the label SDP denotes the parameters in the scheme of
ref.~\cite{DiPierro98}. The other parameters, $B_1^q$ and $\ep_1^q$, are the
same in both schemes. After applying eq.~(\ref{eq:schemi}), the results of
refs.~\cite{DiPierro98,DiPierroproc} read
\bea
&& B^d_1=1.06\pm0.08 \, ,\quad B^d_2=1.01\pm0.07 \, ,\nonumber \\
&& \ep^d_1=-0.01\pm0.03 \, , \quad \ep^d_2=-0.03\pm0.02 \, .
\eea
The differences between these results and those given in table~\ref{tab:inputs}
may give an estimate of the higher orders $1/m_b$ corrections neglected in the
static calculation of ref.~\cite{DiPierro98,DiPierroproc}, and of the
systematic uncertainty introduced by the large extrapolation in the $b$-quark
mass performed in ref.~\cite{APE}. It should be also kept in mind, however,
that the renormalization group evolution has been only performed in
ref.~\cite{DiPierro98,DiPierroproc} with a LO accuracy.

For the $\Lambda_b$ baryon, only the non-perturbative results of an exploratory
study in the HQET
are available at present~\cite{DiPierro:1999tb}. They have been obtained with a
LO accuracy, at a rather large value of the lattice spacing and do not include
the extrapolation of the light quark masses to their physical values. For these
reasons, in quoting the values of the corresponding $B$ parameters, $L_1$ and
$L_2$ in table~\ref{tab:inputs}, we also include in the error our estimate of
the remaining systematic uncertainties. We also note that, in all these
calculations, the matching between the lattice and the continuum theory has
been performed using 1-loop perturbation theory, and that the leading
perturbative corrections are typically found to be large. For this reason, a
non-perturbative evaluation of the relevant renormalization constants would be
very interesting. For $B$ mesons, such a calculation is in
progress~\cite{damir}.

For the large number of $\delta$ and $P$ parameters in
eqs.~(\ref{eq:bpar1})-(\ref{eq:bpar4}), which define the non-valence
contributions and the matrix elements of the penguin operator, only
phenomenological estimates exist~\cite{Chernyak:1995cx,pirjol}. These estimates indicate that
the matrix elements of non-valence operators are suppressed, with
respect to the valence ones, by at least one order of magnitude. These
contributions cancel out in the theoretical evaluation of the lifetime ratio
$\tau(B^+)/\tau(B_d)$, see eq.~(\ref{eq:masterformula}), while they enter the
determination of $\tau(\Lambda_b)/\tau(B_d)$ and $\tau(B_s)/\tau(B_d)$, the
latter for $SU(3)$ breaking effects. Lacking quantitative
calculations of the $\delta$ and $P$ parameters, we will
mainly rely in our numerical analysis on the results of the phenomenological
estimates, and neglect these contributions. A qualitative estimate of the error
introduced by this approximation in the case of the ratio $\tau(\Lambda_b)/
\tau(B_d)$ will be given at the end of the section.

\subsection{Results}
We now present the detailed expressions of the spectator contributions,
represented, in eq.~(\ref{eq:delta}), by the quantities $\Delta_{spec}$. In
these expressions all terms, coming from both valence and non-valence
contributions, which are often omitted in the literature, will be explicitly
taken into account. Using eq.~(\ref{eq:del}), and the definitions
of the $B$-parameters in eqs.~(\ref{eq:bpar1})-(\ref{eq:bpar4}), the spectator
contributions take the following expressions
\bea
\Delta^{B^+}_{\rm\scriptstyle spec}&=& 48\pi^2\, \frac{f_B^2 M_B}{m_b^3 c^{(3)}}
\, \dsum_{k=1}^4 \left(\widetilde c_k^{\,u}-\widetilde c_k^{\,d} \right)
{\cal B}_k^{\, d} \, , \nn \\
\Delta^{B_s}_{\rm\scriptstyle spec}&=&
48\pi^2\, \frac{f_B^2 M_B}{m_b^3 c^{(3)}} \, \left\{ \dsum_{k=1}^4
\left[r \, \widetilde c_k^{\,s} \, {\cal B}_k^{\, s} - \widetilde c_k^{\,d} \,
{\cal B}_k
^{\, d} + \left( \widetilde c_k^{\,u} + \widetilde c_k^{\,d} \right) \left(r\,
\delta_k^{\,ds} - \delta_k^{\,dd} \right) + \right. \right.  \nn \\ && \qquad
\qquad \quad \left. \left.
\widetilde c_k^{\,s} \left(r \, \delta_k^{\,ss} - \delta_k^{\,sd} \right) +
\widetilde c_k^{\,c}
\left(r \, \delta_k^{\,cs} - \delta_k^{\,cd} \right) \right] + c_P
\left(r P^{\, s} - P^{\, d} \right) \right\} \, ,\\
\Delta^{\Lambda}_{\rm\scriptstyle spec}&=&
48\pi^2\, \frac{f_B^2 M_B}{m_b^3 c^{(3)}} \, \left\{ \dsum_{k=1}^4 \left[\left(
 \widetilde c_k^{\,u} + \widetilde c_k^{\,d} \right) {\cal L}_k^{\, \Lambda} -
 \widetilde c_k^{\,d} \, {\cal B}_k^{\, d} + \left( \widetilde c_k^{\,u} +
\widetilde c_k^{\,d} \right) \left(d_k^{\,\Lambda d} - \delta_k^{\,dd} \right)
+ \right. \right.  \nn \\
&& \qquad \qquad \quad \left. \left. \widetilde c_k^{\,s}\left(d_k^{\,
\Lambda s}- \delta_k^{\,sd} \right) + \widetilde c_k^{\,c} \left(d_k^{\,
\Lambda c}- \delta_k^{\,cd} \right) \right] + c_P \left(P^{\,
\Lambda} - P^{\, d} \right)
\right\} \, . \nn
\label{eq:masterformula}
\eea
The $\widetilde c_k$ are the coefficient functions given in
eq.~(\ref{eq:ctilde}). The factor $r$ denotes the ratio $(f_{B_s}^2 M_{B_s})/
(f_B^2 M_B)$ and, in order to simplify the notation, we have defined the
vectors of parameters
\bea
&& \vec {\cal B}^q=\{B_1^q,B_2^q,\ep_1^q,\ep_1^q\} \, ,\nonumber \\
&& \vec {\cal L}=\{L_1,-L_1/2,L_2,-L_2/2\} \, ,\\
&& \vec d^{\Lambda q}=\{\delta^{\Lambda q}_1,-\delta^{\Lambda q}_1/2,
\delta^{\Lambda q}_2,-\delta^{\Lambda q}_2/2\} \nonumber \, .
\eea

Because of the $SU(2)$ symmetry, the non-valence and penguin contributions
cancel out in the expressions of the lifetime ratio $\tau(B^+)/\tau(B_d)$.
From this point of view, the theoretical prediction of this ratio is at
present the most accurate, since it depends only on the non-perturbative
parameters actually computed by current lattice calculations. The prediction of
the ratio $\tau(\Lambda_b)/\tau(B_d)$, instead, is affected by both the
uncertainties on the values of the $\delta$ and $P$ parameters, and by the
unknown expressions of the Wilson coefficients $\widetilde c_k^{\, c}$ and
$c_P$ at the NLO. For the ratio $\tau(B_s)/\tau(B_d)$ the same uncertainties
exists. However they are expected to  be smaller, since these contributions
cancel in the limit of exact $SU(3)$ symmetry.

In order to evaluate the lifetime ratios, we have performed a Monte Carlo
calculation, by extracting the input parameters with flat distributions with
central values and standard deviations given in table \ref{tab:inputs}.
The contributions of all the $\delta$ and $P$ parameters have been neglected.
The strong coupling constant has been kept fixed at the value $\as(m_b)=0.214$.
The values of the bottom and charm quark masses, given in the table and used in
the numerical analysis, correspond to the pole mass definition. The parameter
$c^{(3)}$ in eq.~(\ref{eq:masterformula}) is a function of the ratio
$m_c^2/m_b^2$, and such a dependence has been consistently taken into account
in the numerical analysis. For the range of masses given in table
\ref{tab:inputs},  $c^{(3)}$ varies in the interval $c^{(3)}=3.4\div
4.2$~\cite{gamma4}-\cite{gamma6}.

For convenience, we also present the numerical expression corresponding
to eq.~(\ref{eq:masterformula}), by omitting the non valence and penguin
contributions. We find
\be
\label{eq:magicLO}
\left.
\renewcommand{\arraystretch}{1.2}
\begin{array}{ll}
\Delta^{B^+}_{\rm\scriptstyle spec} \, =&
-\, 0.011(3) \, B_1^d - 0.005(1) \, B_2^d + 0.7(2) \, \ep_1^d -0.19(5)\,\ep_2^d
\,,\\ \\
\Delta^{B_s}_{\rm\scriptstyle spec} \, =&
-\, 0.004(1) \, B_1^s + 0.005(1) \, B_2^s - 0.17(4)\,\ep_1^s + 0.22(6)\,\ep_2^s
\\ &
+\, 0.004(1)\, B_1^d - 0.005(1) \, B_2^d + 0.17(4) \, \ep_1^d -0.19(5)\,\ep_2^d
\,,\\ \\
\Delta^{\Lambda}_{\rm\scriptstyle spec} \, =&
-\, 0.022(6)\, L_1 + 0.24(6)\, L_2 \\ &
+\, 0.004(1) \, B_1^d - 0.005(1)\, B_2^d + 0.17(4)\, \ep_1^d -0.19(5)\, \ep_2^d
\,,\end{array} \,\,\, \right\} \, {\rm LO}
\renewcommand{\arraystretch}{2.0}
\ee
at the LO ($\mu=m_b$), and
\be
\label{eq:magicNLO}
\renewcommand{\arraystretch}{1.2}
\left.
\begin{array}{ll}
\Delta^{B^+}_{\rm\scriptstyle spec} \, =&
-\, 0.07(2) \, B_1^d - 0.011(3) \, B_2^d + 0.7(2) \, \ep_1^d -0.18(5) \, \ep_2^d
\,,\\ \\
\Delta^{B_s}_{\rm\scriptstyle spec} \, =&
-\, 0.007(2) \, B_1^s + 0.009(2) \, B_2^s - 0.16(4)\,\ep_1^s + 0.20(5)\,\ep_2^s
\\ &
+\, 0.007(2)\, B_1^d - 0.008(2) \, B_2^d + 0.16(4) \, \ep_1^d -0.16(4)\, \ep_2^d
\,,\\ \\
\Delta^{\Lambda}_{\rm\scriptstyle spec} \, =&
-\, 0.09(2)\, L_1 + 0.32(8)\, L_2 \\ &
+\, 0.007(2) \, B_1^d - 0.008(2)\, B_2^d + 0.16(4)\, \ep_1^d -0.16(4)\, \ep_2^d
\,,\end{array} \,\,\, \right\} \, {\rm NLO}
\renewcommand{\arraystretch}{2.0}
\ee
at the NLO. Note that we have shown, for each coefficient, an estimate of the
error. These errors, however, are strictly correlated. For this reason,
eqs.~(\ref{eq:magicLO}) and (\ref{eq:magicNLO}) have not been used in the
numerical analysis.
%________________________________________________________________
\begin{figure}
\begin{center}
%\vspace*{-.55cm}
\epsfxsize=14cm
\epsfbox{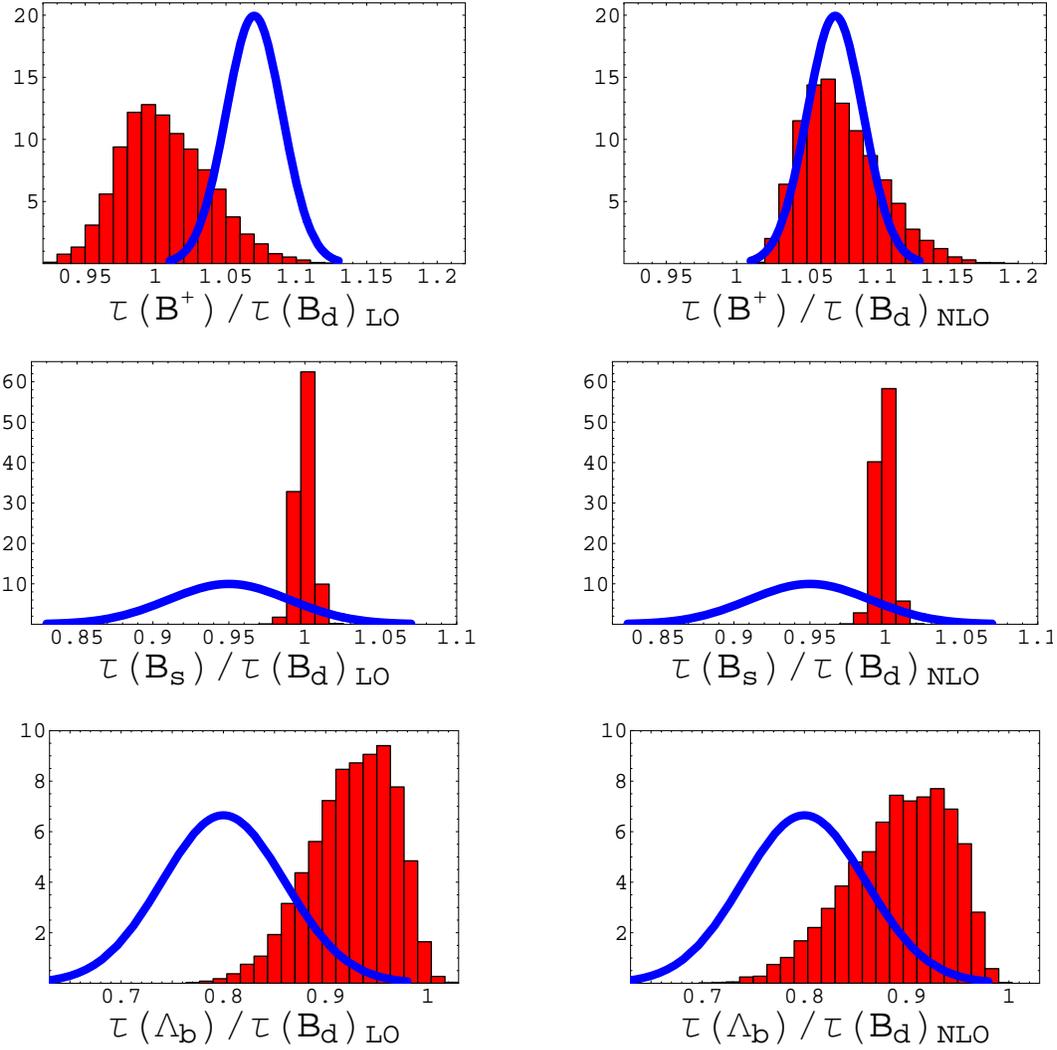}
%\vspace*{-1.3cm}
\end{center}
\caption{\it Theoretical (histogram) vs experimental (solid line) distributions of
lifetime ratios.
The theoretical predictions are shown at the LO (left) and NLO (right).}
\label{fig:plot}
\end{figure}
%________________________________________________________________

We finally present our predictions for the lifetimes ratios of beauty hadrons,
which have also been quoted in the introduction. Using the LO expressions of
the Wilson coefficients, we obtain
\be
\label{eq:reslo}
\left. \frac{\tau(B^+)}{\tau(B_d)}\, \right|_{\rm LO}\!\!\!\!\! =
1.01 \pm 0.03 \, ,  \qquad
\left. \frac{\tau(B_s)}{\tau(B_d)}\, \right|_{\rm LO}\!\!\!\!\! =
1.00 \pm 0.01 \, ,  \qquad
\left. \frac{\tau(\Lambda_b)}{\tau(B_d)}\, \right|_{\rm LO}\!\!\!\!\! =
0.93 \pm 0.04\, ,
\ee
while, including the NLO corrections computed in this paper, we get
\be
\label{eq:resnlo}
\left. \frac{\tau(B^+)}{\tau(B_d)}\, \right|_{\rm NLO}\!\!\!\!\! =
1.07 \pm 0.03 \, ,  \qquad
\left. \frac{\tau(B_s)}{\tau(B_d)}\, \right|_{\rm NLO}\!\!\!\!\! =
1.00 \pm 0.01 \, ,  \qquad
\left. \frac{\tau(\Lambda_b)}{\tau(B_d)}\, \right|_{\rm NLO}\!\!\!\!\! =
0.89 \pm 0.05\, .
\ee
The central values and errors quoted in eqs.~(\ref{eq:reslo})
and~(\ref{eq:resnlo}) are the average and the standard deviation of the
theoretical distributions. These distributions are shown in fig.~\ref{fig:plot},
together with the experimental ones.

In the LO results, the errors include the effect of varying the renomalization
scale $\mu$ between $m_b/2$ and $2 m_b$.
At the NLO, the variation with $\mu$ of the Wilson coefficients is compensated by the
$\mu$ dependence of the renormalized operators, up to NNLO corrections.
We cannot estimate the residual scale dependence because we only know the non-perturbative
the relevant operators at a single scale $\mu=m_b$. Therefore the renormalization scale is kept
fixed in the NLO analysis.

We find that, with the inclusion of the NLO
corrections, the theoretical prediction for the ratio
$\tau(B^+)/\tau(B_d)$ turns out to be in very good agreement with the
experimental measurement, given in eq.~(\ref{eq:rexp}). For the ratios
$\tau(B_s)/\tau(B_d)$ and $\tau(\Lambda_b)/\tau(B_d)$ the agreement is also
very satisfactory, and the difference between theoretical and experimental
determinations is at the $1\sigma$ level.

The errors quoted in eq.~(\ref{eq:resnlo}) do not take into account the systematic uncertainty
due to terms of ${\cal O}(\alpha_s m_c^2/m_b^2)\sim 0.1\,\alpha_s$, not included in our
calculation. Na\"{\i}vely, these corrections can change
the NLO terms by about $10\%$, thus affecting the lifetime ratios at the level of $1\%$. A realistic
estimate, however, requires an explicit calculation, which is under way.

We conclude this analysis with a qualitative estimate of the contribution
of the non valence and penguin matrix elements, which have been neglected in
the  evaluation of $\tau(B_s)/\tau(B_d)$ and $\tau(\Lambda_b)/\tau(B_d)$. 
We consider as an example the term
\be
\label{eq:ultima}
48\pi^2\, \frac{f_B^2 M_B}{m_b^3 c^{(3)}} \, \dsum_{k=1}^4 \left[
\left( \widetilde c_k^{\,u} + \widetilde c_k^{\,d} \right) \left(
d_k^{\,\Lambda d} - \delta_k^{\,dd} \right) \right]
\ee
entering the expression of $\Delta^{\Lambda}_{\rm\scriptstyle spec}$ 
in eq.~(\ref{eq:masterformula}). Since the coefficient functions
$\widetilde c_k^{\,u}$ and $\widetilde c_k^{\,d}$ are known at the NLO, the
only uncertainty in eq.~(\ref{eq:ultima}) is the value of the non-perturbative
parameters. According to the phenomenological estimates of ref.~\cite{pirjol},
the typical values of these parameters are smaller by approximately one order
of magnitude with respect to the matrix elements of the valence operators.
By choosing for all the
differences $d_k^{\,\Lambda d} - \delta_k^{\,dd}$ a common value of 0.05, we
then find that the ratio $\tau(\Lambda_b)/\tau(B_d)$ varies by less than 3\%, 
and the variation is approximately proportional to the values of the $\delta$ 
parameters. The matrix elements of the penguin operators are not expected to be
smaller than those of the valence operators. Numerically, since the 
coefficient function $c_P$ vanishes at the LO, this contribution is expected
to have the size of a typical NLO corrections. It is clear that a quantitative
evaluation of the non-valence and penguin operators would be very interesting,
in order to further improve the agreement between theoretical and experimental
determinations of the lifetime ratios of beauty hadrons.

%%%%%%%%%%%%%%%%%%%%%%%%%%%%%%%%%%%%%%%%%%%%%%%%%%%%%%%%%%%%
\section*{Acknowledgments}
%%%%%%%%%%%%%%%%%%%%%%%%%%%%%%%%%%%%%%%%%%%%%%%%%%%%%%%%%%%%
We are grateful to U. Aglietti, D. Becirevic, G. Martinelli, J. Reyes and C.T. Sachrajda
for interesting  discussions on the subject of this paper. We thank M. Di Pierro
for correspondence on the lattice results of
refs.~\cite{DiPierro98}-\cite{DiPierroproc}. M.C. and F.M. thank the TH
division at CERN where part of this work has been done. Work  partially supported by
the European  Community's Human Potential Programme under HPRN-CT-2000-00145
Hadrons/Lattice QCD.

%%%%%%%%%%%%%%%%%%%%%%%%%%%%%%%%%%%%%%%%%%%%%%%%%%%%%%%%%%%%
\section*{Appendix}
%%%%%%%%%%%%%%%%%%%%%%%%%%%%%%%%%%%%%%%%%%%%%%%%%%%%%%%%%%%%
In this appendix we discuss in some details the r\^ole of the evanescent
operators in the matching procedure. Because of the IR divergences, the matrix
elements of renormalized evanescent operators do not vanish in the $D\to 4$
limit. Therefore, these operators contribute, at the NLO, to the matching of
the physical operators.

For the sake of simplicity, we consider here the abelian case. The inclusion of
colour factors is trivial.

Let us consider the calculation of the imaginary part of the double insertion
of the operator
\be
Q=(\bar b c )_{V-A} (\bar u d)_{V-A}
\ee
in the left diagram of fig.~\ref{fig:lospec}. Neglecting the charm quark
mass, the contribution to the amplitude in the full theory, proportional to the
operators with the flavour structure $\bar b b \bar d d$, is
\bea
{\cal T}^{(0)} & = &
-\dfrac{G_F^2 \vert V_{cb}\vert^2 m_b^2}{2\pi}C(\mu)^2
\Bigg\{\left[ \frac{1}{24} + \frac{\varepsilon}{9} +
\frac{1}{24} \varepsilon \log\dfrac{\mu^2}{m_b^2} \right] \,
(\bar b \gamma^\mu \gamma^\alpha \gamma^\nu_L b) (\bar d
\gamma_\nu \gamma_\alpha \gamma_{\mu L} d) + \nn \\
&& \left[ \frac{1}{12} + \frac{5}{36}\varepsilon +
\frac{1}{12} \varepsilon \log\dfrac{\mu^2}{m_b^2} \right] \,
\frac{1}{m_b^2} (\bar b \gamma^\mu \slash{p}_b \gamma^\nu_L b) (\bar d
\gamma_\nu \slash{p}_b \gamma{\mu L} d)\Bigg\}\, ,
\eea
where $C(\mu)$ is the $\DB =1$ Wilson coefficient of $Q$ and
$\gamma^\mu_{R,L}=\gamma^\mu (1 \pm \gamma_5)$.
Using the equations of motion, the result can be written as
\bea
\label{eq:tq2}
{\cal T}^{(0)} =
-\dfrac{G_F^2 \vert V_{cb}\vert^2 m_b^2}{2\pi} &&\!\!\!\!\! C(\mu)^2
\Bigg\{\left[ \frac{1}{9} + \frac{5}{27}\varepsilon +
\frac{1}{9} \varepsilon \log\dfrac{\mu^2}{m_b^2} \right] \,
 (\bar b \gamma^\mu_L b) (\bar d \gamma_{\mu L} d) - \nn \\
&& \left[ \frac{1}{18} + \frac{5}{54}\varepsilon +
\frac{1}{18} \varepsilon \log\dfrac{\mu^2}{m_b^2} \right] \,
 (\bar b \gamma^\mu_R b) (\bar d \gamma_{\mu L} d) +  \\
&& \left[ \frac{1}{18} + \frac{29}{216}\varepsilon +
\frac{1}{18} \varepsilon \log\dfrac{\mu^2}{m_b^2} \right] \,
 (\bar b \gamma^\mu \gamma^\alpha \gamma^\nu_L b) (\bar d
\gamma_\nu \gamma_\alpha \gamma_{\mu L} d) + \nn\\
&& \left[ \frac{1}{72} + \frac{5}{216}\varepsilon +
\frac{1}{72} \varepsilon \log\dfrac{\mu^2}{m_b^2} \right] \,
 (\bar b \gamma^\mu \gamma^\alpha \gamma^\nu_R b) (\bar d
\gamma_\nu \gamma_\alpha \gamma_{\mu L} d)\Bigg\}\,.\nn
\eea
We then introduce two physical operators and two evanescent operators
(in the non-abelian case there are four of them, see eqs.~(\ref{eq:opeff})
and~(\ref{eq:ev})) defined as~\footnote{We explicitly checked that alternative
definitions of evanescent operators, which differ from those in
eq.~(\ref{eq:eva}) by terms of ${\cal O}(\varepsilon)$, give the
same results in the final matching.}
\be
\label{eq:eva}
\begin{array}{ll}
O_1=\bar b \gamma^\mu_L b\; \bar d \gamma_{\mu L} d \, , &
E_{1} = \bar b \gamma_\mu \gamma_\alpha \gamma_{\nu\, L} b\; \bar d
\gamma^\nu \gamma^\alpha \gamma^\mu_L d -
4 \, O_1 \, , \\
O_2=\bar b \gamma^\mu_R b\; \bar d \gamma_{\mu L} d \, , &
E_{2} = \bar b \gamma_\mu \gamma_\alpha \gamma_{\nu\, R} b\; \bar d
\gamma^\nu \gamma^\alpha \gamma^\mu_L d -16 \, O_2 \, .
\end{array}
\ee
Using eqs.~(\ref{eq:tq2}) and~(\ref{eq:eva}), the LO matching is easily
performed and gives
\be
\widehat \Gamma^{(0)}_{\rm\scriptstyle spect} = \frac{G_F^2 \vert V_{cb}\vert^2
m_b^2}{2\pi} C(\mu)^2 \dsum_{k=1,2}\left( c_{O_k} \, O_k + c_{E_k} \,
E_k\right)\,,
\ee
where the coefficient functions at the LO, including terms of 
${\cal O}(\varepsilon)$, are
\be
\label{eq:c0eps}
\begin{array}{ll}
c_{O_1} = -\dfrac{1}{3} - \dfrac{13}{18}\varepsilon -
\dfrac{1}{3} \varepsilon \log\dfrac{\mu^2}{m_b^2}  \, , \quad &
c_{E_1} = -\dfrac{1}{18} \,,\\
c_{O_2} = -\dfrac{1}{6} - \dfrac{5}{18}\varepsilon -
\dfrac{1}{6} \varepsilon \log\dfrac{\mu^2}{m_b^2} \,,
& c_{E_2} = -\dfrac{1}{72}\,.
\end{array}
\ee
The renormalized evanescent operators $E_k$ have to be inserted in the
effective theory at the NLO.  Due to the presence of IR poles, they acquire
non-vanishing matrix elements in the $D\to 4$ limit. These finite contributions 
enter the matrix $\hat s$ in eq.~(\ref{eq:Texp}) and, according to 
eq.~(\ref{eq:match1}), contribute to the final determination of the Wilson 
coefficients at the NLO.

A finite contribution is also obtained from the terms of ${\cal O}(\varepsilon)$
in the coefficient functions $c_{O_k}$ of the physical operators. Indeed, in
eq.~(\ref{eq:match1}), these coefficients multiply the matrix $\hat s$ which, 
because of the IR divergences, contains poles in $1/\varepsilon$. Note that 
these terms also depend on the renormalization scale $\mu$ and are needed 
for reconstructing the proper UV scale dependence of the Wilson coefficients at 
the NLO.

%%%%%%%%%%%%%%%%%%%%%%%%%%%%%%%%%%%%%%%%
% thebibliography
%%%%%%%%%%%%%%%%%%%%%%%%%%%%%%%%%%%%%%%%

\end{document}